\documentclass[preprint,aps,nofootinbib]{revtex4-1}
\pdfoutput=1
\usepackage{bm, amsmath,amssymb,amsfonts,slashed,array,graphicx,soul}
\usepackage[vcentermath,noautoscale]{youngtab}
\usepackage{mathrsfs}
\usepackage{xcolor}
\usepackage{hyperref}
\usepackage{cancel}
\usepackage{booktabs}
\usepackage{multirow}
\usepackage{enumitem}
\usepackage[normalem]{ulem}
\usepackage{slashed}            % for slashed characters in math mode
\usepackage{tikz}
\usetikzlibrary{arrows,shapes}
\usetikzlibrary{trees}
\usetikzlibrary{matrix,arrows} 				% For commutative diagram
\usetikzlibrary{positioning}				% For "above of=" commands
\usetikzlibrary{calc,through}				% For coordinates
\usetikzlibrary{decorations.pathreplacing}  % For curly braces
\usetikzlibrary{decorations.pathmorphing}	% For Feynman Diagrams
\usetikzlibrary{decorations.markings}
\usetikzlibrary{snakes}
\usepackage[normalem]{ulem}
\tikzset{
    vector/.style={decorate, decoration={snake}, draw},
	provector/.style={decorate, decoration={snake,amplitude=2.5pt}, draw},
	antivector/.style={decorate, decoration={snake,amplitude=-2.5pt}, draw},
    fermion/.style={draw=black, postaction={decorate},
        decoration={markings,mark=at position .55 with {\arrow[draw=black]{>}}}},
    fermionbar/.style={draw=black, postaction={decorate},
        decoration={markings,mark=at position .55 with {\arrow[draw=black]{<}}}},
    fermionnoarrow/.style={draw=black},
    gluon/.style={decorate, draw=black,
        decoration={coil,amplitude=4pt, segment length=5pt}},
    scalar/.style={dashed,draw=black, postaction={decorate},
        decoration={markings,mark=at position .55 with {\arrow[draw=black]{>}}}},
    scalarbar/.style={dashed,draw=black, postaction={decorate},
        decoration={markings,mark=at position .55 with {\arrow[draw=black]{<}}}},
    scalarnoarrow/.style={dashed,draw=black},
    electron/.style={draw=black, postaction={decorate},
        decoration={markings,mark=at position .55 with {\arrow[draw=black]{>}}}},
	bigvector/.style={decorate, decoration={snake,amplitude=4pt}, draw},
}

\def\lsim{\mathrel{\rlap{\lower4pt\hbox{\hskip1pt$\sim$}}
    \raise1pt\hbox{$<$}}}
\def\gsim{\mathrel{\rlap{\lower4pt\hbox{\hskip1pt$\sim$}}
    \raise1pt\hbox{$>$}}}

\renewcommand{\thefootnote}{\fnsymbol{footnote}}
\begin{document}
\title{Dark matter shifts away from direct detection}
\author{Reuven Balkin, Maximilian Ruhdorfer, Ennio Salvioni and Andreas Weiler }
\email[Email: ]{reuven.balkin@tum.de}\email{max.ruhdorfer@tum.de}\email{ennio.salvioni@tum.de}\email{andreas.weiler@tum.de}
\affiliation{Physik-Department, Technische Universit\"at M\"unchen, 85748 Garching, Germany}
%\date{\today}
\begin{abstract}

\noindent Dark matter could emerge along with the Higgs as a composite pseudo-Nambu-Goldstone boson $\chi$ with decay constant $f\sim \mathrm{TeV}$. This type of WIMP is especially compelling because its leading interaction with the Standard Model, the {\it derivative Higgs portal}, has the correct annihilation strength for thermal freeze-out if $m_\chi \sim O(100)$ GeV, but is negligible in direct detection experiments due to the very small momentum transfer. The explicit breaking of the shift symmetry which radiatively generates $m_\chi$, however, introduces non-derivative DM interactions. In existing realizations a marginal Higgs portal coupling $\lambda$ is generated with size comparable to the Higgs quartic, and thus well within reach of XENON1T. Here, we present and analyze the interesting case where the pattern of explicit symmetry breaking naturally suppresses $\lambda$ beyond the reach of current and future direct detection experiments. If the DM acquires mass from bottom quark loops, the bottom quark also mediates suppressed DM--nucleus scattering with cross sections that will be eventually probed by LZ. Alternatively, the DM can obtain mass from gauging its stabilizing $U(1)$ symmetry. No direct detection signal is expected even at future facilities, but the introduction of a dark photon $\gamma_D$ has a number of phenomenological implications which we study in detail, treating $m_{\gamma_D}$ as a free parameter. Complementary probes of the dark sector include indirect DM detection, DM self-interactions, and extra radiation, as well as collider experiments. We frame our discussion in an effective field theory, motivating our parameter choices with a detailed analysis of an $SO(7)/SO(6)$ composite Higgs model, which can yield either scenario at low energies.

\end{abstract}
\preprint{TUM-HEP-1162-18}
\maketitle
\tableofcontents
%%%%%%%%%%%%%%%%%%%%%%%
\renewcommand{\thefootnote}{\arabic{footnote}}
\section{Introduction}
\noindent The idea that the Higgs may arise as a composite pseudo Nambu-Goldstone boson (pNGB) from a new strongly-interacting sector provides one of the best-motivated solutions to the naturalness problem of the weak scale~\cite{Panico:2015jxa,Bellazzini:2014yua}. The minimal realistic model \cite{Agashe:2004rs} is based on a strong sector with global symmetry $\mathcal{G} = SO(5)$, whose spontaneous breaking to $\mathcal{H} = SO(4)$ at scale $f \sim \mathrm{TeV}$ yields four real pNGBs, identified with the components of the Standard Model (SM)-like Higgs doublet $H$. 

Non-minimal models offer, in addition, an appealing link to the dark matter (DM) puzzle~\cite{Frigerio:2012uc}. If one of the extra pNGBs contained in $\mathcal{G}/\mathcal{H}$, assumed to be a SM singlet and labeled $\chi$, is stable (owing, for example, to a discrete $Z_2$ symmetry~\cite{Frigerio:2012uc}, or because it is the lightest particle charged under a continuous $U(1)$ symmetry~\cite{Balkin:2017aep}), it constitutes a compelling weakly interacting massive particle (WIMP) DM candidate. Not only is $\chi$ naturally light and weakly coupled in the same way the Higgs is, but also its leading coupling to the SM is the {\it derivative Higgs portal}, 
\begin{equation}
\frac{1}{f^2}\partial_\mu |H|^2 \partial^\mu |\chi|^2 \,,
\end{equation}
which is extremely suppressed at the small momentum transfers that characterize DM scattering with heavy nuclei, $|t| / f^2 \lesssim (100\;\mathrm{MeV})^2/ (1 \;\mathrm{TeV})^2 \sim 10^{-8}$. This type of WIMP is therefore naturally compatible with the existing strong exclusions from direct detection experiments. At the same time, the interaction strength of DM annihilation is $s / f^2 \simeq 4 m_\chi^2 / f^2$, which if the DM acquires a radiative mass $m_\chi \sim 100\;\mathrm{GeV}$ is in the right range to obtain the observed relic density through thermal freeze-out.

This simple and attractive picture can, however, be significantly altered by explicit symmetry breaking effects. Some sources of explicit breaking of the Goldstone shift symmetries are in fact necessary, in order to provide a potential and Yukawa couplings for $H$ and at least the mass for $\chi$. Generically, these sources also introduce non-derivative couplings between the DM and the SM, in particular the {\it marginal Higgs portal}, 
\begin{equation}
\lambda |H|^2 |\chi|^2  \,,
\end{equation}
which is strongly constrained by direct detection. The main purpose of this paper is to construct and analyze realistic models where $\lambda$ is very suppressed, either because it is proportional to the Yukawas of the light SM fermions or because it arises at higher-loop order, while at the same time $\chi$ obtains a mass of $O(100)$ GeV at one loop. These models then retain the most appealing features of pNGB DM discussed above, and should in our view be considered as very motivated targets for experiments that will search for WIMPs in the near future.

The irreducible sources of explicit $\mathcal{G}$ breaking are the gauging of $SU(2)_L\times U(1)_Y \subset \mathcal{H}$, and the couplings of the SM fermions. If $\chi$ is a SM-singlet, as assumed in this paper, the SM gauging does not give it a potential at one loop. The fermions are linearly coupled to operators of the strong sector, thus realizing the partial compositeness mechanism \cite{Kaplan:1991dc}. Hence we must specify the incomplete $\mathcal{G}$ representations (spurions) that $q_L = (t_L, b_L)^T, t_R$ and $b_R$ are embedded into, where we focus on the third-generation quarks since the first two generations of quarks and the leptons have much smaller Yukawa couplings. The choice of the spurions fixes the explicit breaking of the DM shift symmetry and therefore the strength of the non-derivative couplings of $\chi$. Three qualitatively different scenarios can be identified:

\vspace{1mm}
\noindent {\bf Leading breaking by top quark couplings.} This case was first discussed in Ref.~\cite{Frigerio:2012uc} and later analyzed extensively in the $SO(6)/SO(5)$ model, where the DM is a real scalar stabilized by a $Z_2$ symmetry~\cite{Marzocca:2014msa}, and by the authors of this paper in the $SO(7)/SO(6)$ model, where the DM is a complex pNGB stabilized by a $U(1)$~\cite{Balkin:2017aep}. Top loops make the DM heavier than the Higgs, and the global symmetry causes the marginal portal coupling to be generated with size comparable to that of the Higgs quartic $\lambda_h\,$,
\begin{equation}
\lambda \lesssim \frac{\lambda_h}{2} \qquad\; \text{and} \qquad  m_\chi \gg m_h\,.
\end{equation}
As a consequence, once the observed $\lambda_h \simeq 0.13$ is reproduced, $\lambda$ is automatically of few percent, corresponding to DM-nucleon cross sections $\sigma_{\rm SI}^{\chi N} \sim 10^{-46}\,\mathrm{cm}^2$ that are currently being probed by XENON1T. This setup does not fully realize the pNGB DM picture in the sense described above and has been well covered in previous work, so it will not be studied further in this paper. Nevertheless, for completeness we provide a very short summary of its features in Sec.~\ref{eq:tRbreaking}.

\vspace{1mm}
\noindent {\bf Leading breaking by bottom quark couplings.} Although this case was also first discussed in Ref.~\cite{Frigerio:2012uc}, here we focus on a different parametric regime. The marginal portal coupling and DM mass are both generated at one loop, but scale very differently with the bottom Yukawa, 
\begin{equation} \label{eq:bottomScalings}
\lambda \propto y_b^2 \qquad\; \text{and} \qquad m_\chi \propto (y_b g_{\ast})^{1/2} f\,,
\end{equation}
where $g_\ast$ is the strong sector coupling. Hence $\lambda$ is so small ($\lambda \ll 10^{-3}$) that it is irrelevant for direct detection, but $\chi$ can be sufficiently heavy ($m_\chi \sim 100\;\mathrm{GeV}$) that its annihilation via the derivative Higgs portal yields the correct DM relic density. The explicit breaking of the $\chi$ shift symmetry, however, also generates the operator $y_b \bar{q}_L H b_R\, |\chi|^2/f^2$, which yields small DM-nucleon cross sections $\sigma_{\rm SI}^{\chi N} \sim  10^{-47}\,\mathrm{cm}^2$. Part of the parameter space will therefore be tested in next-generation direct detection experiments such as LZ. This setup is discussed in Sec.~\ref{eq:bRbreaking}, with further important details provided in Appendix~\ref{sec:appA}.

\vspace{1mm}
\noindent {\bf DM shift symmetry unbroken by SM fermion couplings.} It is possible to embed all SM quarks into spurions that preserve the DM shift symmetry \cite{Frigerio:2012uc,Balkin:2017aep}. In this case some beyond-the-SM source of explicit breaking is required, in order to generate a mass for the DM. In Sec.~\ref{sec:gaugeBreaking}, which contains the main results of this paper, we focus on the case where $\chi$ is a complex scalar stabilized by a $U(1)_{\rm DM}\subset \mathcal{H}$ symmetry, and show that gauging $U(1)_{\rm DM}$ with coupling $g_D$ can naturally produce a one-loop mass 
\begin{equation}
m_\chi \propto g_D f
\end{equation}
of $O(100)$ GeV for $\chi$, while $\lambda$ is strongly suppressed as it is only generated at higher loop order. This setup realizes the crucial feature of the previously-advertised pNGB DM picture, namely the DM scattering on nuclei is too suppressed to be within the foreseeable reach of direct detection experiments. Incidentally, let us mention that gauging $U(1)_{\rm DM}$ may also increase the theoretical robustness of the DM stability, an aspect that will be briefly addressed in our final remarks of Sec.~\ref{sec:outlook}. 

The presence of a dark photon $\gamma_D$ yields a rich phenomenology, which we study in detail. Importantly, we take zero kinetic mixing between the $U(1)_{\rm DM}$ and the SM hypercharge. This is motivated by our explicit analysis of the $SO(7)/SO(6)$ model, where the kinetic mixing is forbidden because $C_D$, the charge conjugation associated to $U(1)_{\rm DM}$, is an accidental symmetry of the theory, at least in the limit where subleading spurions for the SM fermions are neglected. This point is thoroughly discussed in Appendix~\ref{app:KinMixing}. First we examine, in Sec.~\ref{subsec:masslessDP}, the case where the dark photon is massless and therefore constitutes dark radiation. The dark sector phenomenology shares several aspects with those considered in Refs.~\cite{Feng:2008mu,Ackerman:mha,Feng:2009mn,Agrawal:2016quu},\footnote{See also the scenario of Refs.~\cite{Fan:2013yva,Fan:2013tia}, where only a subdominant component of the DM is assumed to be charged under a hidden $U(1)$, and in addition features dissipative dynamics.} predicting an array of signals in cosmology, astroparticle, and collider experiments. These signatures place constraints on the parameter space and will allow this scenario to be further probed in the near future. Subsequently, we consider in Sec.~\ref{subsec:massiveDP} the possibility that $\gamma_D$ acquires a mass through the St\"uckelberg mechanism. Here again $C_D$ invariance plays an important role, making $\gamma_D$ stable for $m_{\gamma_D} < 2m_\chi$, when the decay to $\chi \chi^\ast$ is kinematically forbidden. We identify a region of parameters where both $\chi$ and $\gamma_D$ behave as cold DM, and discuss the novel features of this two-component-DM regime.

\vspace{1mm}
Our discussion is phrased within a low-energy effective field theory (EFT), but as already stated we support our parameter choices with concrete examples that arise in the composite Higgs model based on the $SO(7)/SO(6)$ symmetry breaking pattern \cite{Balkin:2017aep}. This construction can yield each of the three above scenarios depending on the region of parameters one focuses on, and is therefore well suited as theory backdrop. Other previous work on composite Higgs models with pNGB DM includes Refs.~\cite{Chala:2012af,Barnard:2014tla,Kim:2016jbz,Chala:2016ykx,Ma:2017vzm,Ballesteros:2017xeg,Balkin:2017yns,Alanne:2018xli}, whereas Refs.~\cite{Fonseca:2015gva,Brivio:2015kia,Bruggisser:2016ixa,Bruggisser:2016nzw} performed studies employing EFTs. 

The remainder of this paper is organized as follows. In Sec.~\ref{sec:effLagr} we introduce the EFT we use to describe the pNGB Higgs and DM, as well as its essential phenomenological implications. Section~\ref{sec:fermionbreak} presents concisely the scenarios where the DM shift symmetry is explicitly broken by the couplings of the SM fermions. In Sec.~\ref{sec:gaugeBreaking} we analyze in depth the case where the DM shift symmetry is preserved by the couplings of the SM fermions, but broken by the gauging of the $U(1)$ symmetry that stabilizes the DM. Finally, Sec.~\ref{sec:outlook} provides some closing remarks. Appendices~\ref{sec:appA}, \ref{sec:appB} and \ref{app:KinMixing} contain details on important aspects of the $SO(7)/SO(6)$ composite Higgs model, while Appendix~\ref{app:PhenoResults} collects formulas relevant to our phenomenological analysis.

%%%%%%%%%%%%%%%
\section{Effective Lagrangian for the Higgs and DM pNGBs} \label{sec:effLagr}
\noindent The low-energy effective Lagrangian for the pNGBs, namely the Higgs doublet $H$ and the SM-singlet DM, taken to be a complex scalar $\chi$ stabilized by a $U(1)_{\rm DM}$ symmetry,\footnote{For real DM $\eta$ that is stable due to a $Z_2$ symmetry, we simply replace $\chi \to \eta /\sqrt{2}$ in $\mathcal{L}_{\rm eff}$.} has the form
\begin{equation}
\mathcal{L}_{\rm eff} = \mathcal{L}_{\rm GB} + \mathcal{L}_f - V_{\rm eff}\,,
\end{equation}
where $\mathcal{L}_{\rm GB}$ contains only derivative interactions, whose structure is determined by the non-linearly realized global symmetry. $\mathcal{L}_f$ contains the couplings to the SM fermions, which originate from elementary-composite mixing couplings that break $\mathcal{G}$ explicitly. These elementary-composite mixings, together with the gauging of a subgroup of $\mathcal{G}$ that includes the SM electroweak symmetry, generate the radiative potential $V_{\rm eff}$. We discuss first the leading order Lagrangian $\mathcal{L}_{\rm GB}$, and then turn to the effects of the explicit symmetry breaking, contained in $\mathcal{L}_{f} - V_{\rm eff}$.

\subsection{Two-derivative Lagrangian}\label{subsec:LGB}
The most general two-derivative, $SU(2)_L\times SU(2)_R \times U(1)_{\rm DM}\subset \mathcal{H}$ invariant Lagrangian\footnote{More precisely, this is the most general $SU(2)_L\times SU(2)_R \times U(1)_{\rm DM}$ invariant Lagrangian where $SU(2)_R$ is only broken by the gauging of hypercharge.\vspace{1.5mm}} that arises from the nonlinear sigma model kinetic term is\footnote{We do not include in $\mathcal{L}_{\rm GB}$ operators containing $\chi^*\overset\leftrightarrow{\partial_\mu}\chi \equiv  \chi^\ast \partial_\mu \chi - \partial_\mu \chi^\ast \chi$, which vanish trivially in the $SO(6)/SO(5)$ model where $\chi \to \eta/\sqrt{2}$ with real $\eta$, and are forbidden in the $SO(7)/SO(6)$ model by custodial $SO(4) \simeq SU(2)_L \times SU(2)_R$ invariance, since $H$ and $\chi$ belong to the same irreducible representation of $\mathcal{H} = SO(6)$. Notice also that $\chi^*\overset\leftrightarrow{\partial_\mu}\chi$ is odd under the charge conjugation associated to $U(1)_{\rm DM}$.}
\begin{equation}\label{eq:LGB}
\mathcal{L}_{\rm GB} \,=\, |D^\mu H|^2 + |\partial^\mu \chi|^2  +  \frac{c_H}{2f^2} \partial_\mu |H|^2 \partial^\mu |H|^2 + \frac{c_d}{f^2} \partial_\mu |H|^2 \partial^\mu |\chi |^2 + \frac{c_\chi}{2f^2} \partial_\mu |\chi |^2 \partial^\mu |\chi |^2 . %+ \frac{\tilde{c}_\chi}{f^2} (\chi^*\overset\leftrightarrow{\partial_\mu}\chi)^2 ,
\end{equation}
We could have written four additional operators, 
\begin{equation}
\frac{c_1}{f^2} |D_\mu H|^2  |H|^2 \,,\qquad \frac{c_2}{f^2}  |D_\mu H |^2 |\chi |^2 \,, \qquad \frac{c_3}{f^2} |\partial_\mu \chi |^2  |H|^2 \,,\qquad \frac{c_4}{f^2}  |\partial_\mu \chi |^2 |\chi |^2 \,,
\end{equation}
but these can be removed through the $O(1/f^2)$ field redefinition
\begin{equation}
\label{eq:FieldRedef}
H\rightarrow \Big(1  - \frac{c_1}{2f^2} |H|^2 - \frac{c_2}{2 f^2} |\chi |^2  \Big) H \,,\qquad \chi\rightarrow \Big( 1 - \frac{c_3}{2 f^2} |H|^2 -\frac{c_4}{2f^2} |\chi|^2 \Big) \chi \, .
\end{equation}
Notice that for $c_1=c_2=c_3=c_4=-2/3$ these are the leading terms of
\begin{equation}
\frac{\sin (\pi /f )}{\pi}\, \pi^a \rightarrow \frac{\pi^a}{f}\,,\qquad \pi = \sqrt{\vec{\pi}^{\,T}\vec{\pi}}\,,
\end{equation} 
where $\vec{\pi}$ is the GB vector~\cite{Gripaios:2009pe}. This redefinition has customarily been adopted in studies of the $SO(6)/SO(5)$ and $SO(7)/SO(6)$ models because in the basis of Eq.~(\ref{eq:LGB}), which also coincides with the SILH basis \cite{Giudice:2007fh} when restricted to Higgs interactions, the scalar potential is a simple polynomial and the vacuum expectation value (VEV) of the Higgs is equal to $v \simeq 246\;\mathrm{GeV}$. In those models the coefficients take the values $c_H = c_d = c_\chi = 1$, which we often adopt as reference in the following.

\begin{figure}[t]
\centering
\includegraphics[width=.495\textwidth]{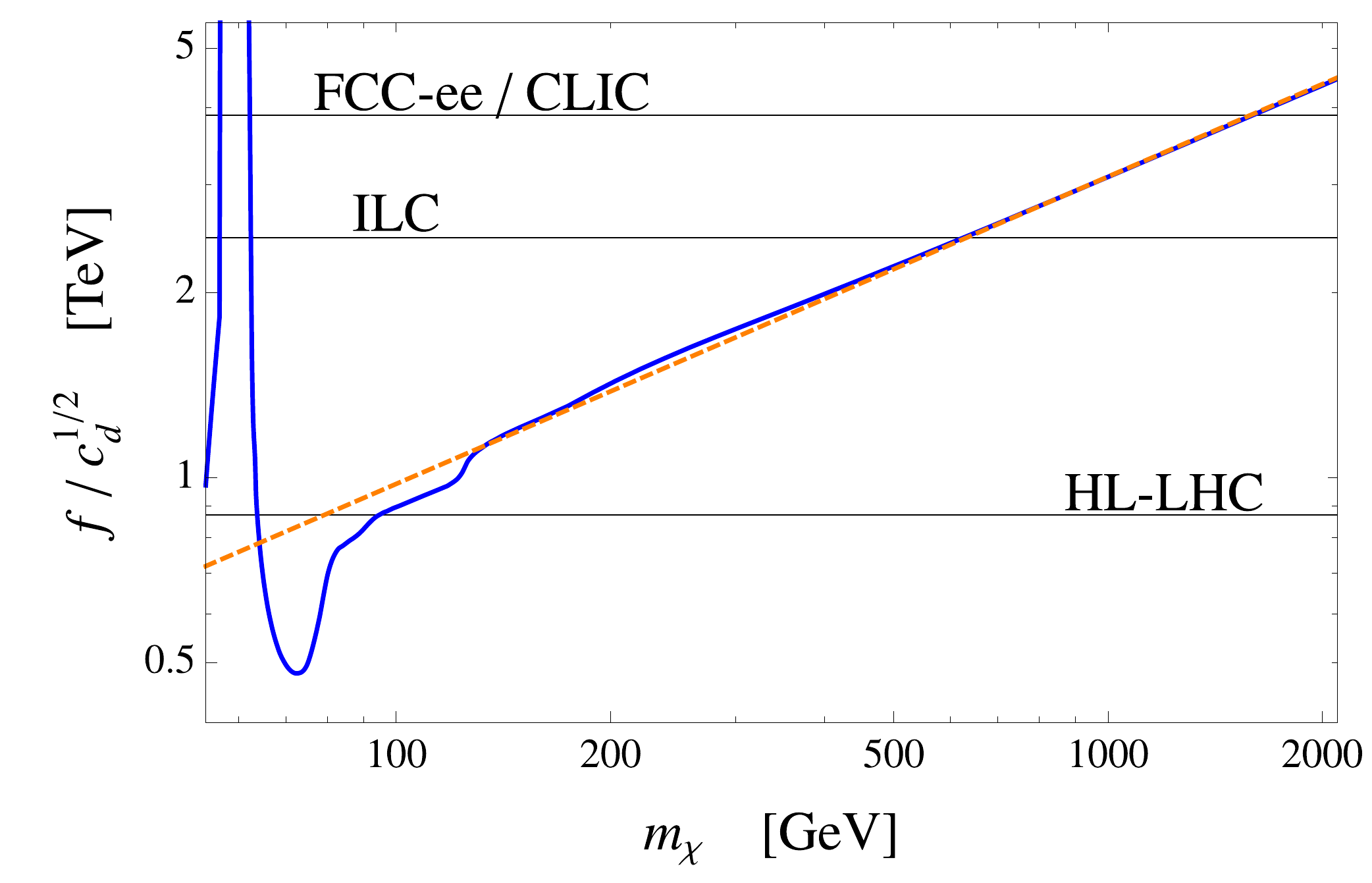}
\includegraphics[width=.495\textwidth]{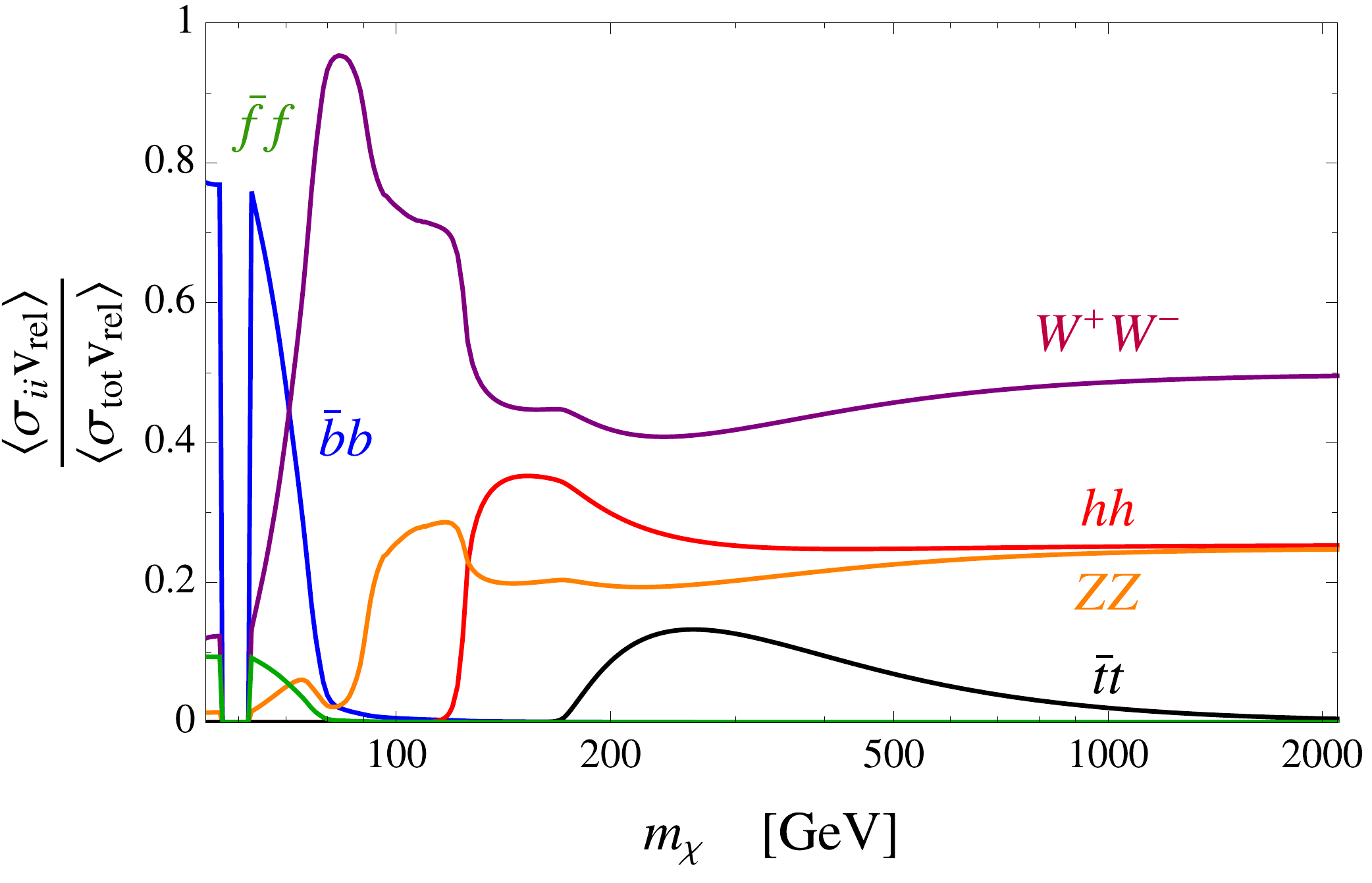}
\caption{{\it Left panel:} value of the global symmetry breaking scale $f$ that allows to reproduce the observed DM relic density via the derivative Higgs portal, as function of the DM mass. In solid blue the full Boltzmann solution, in dashed orange the approximate relation given in Eq.~\eqref{eq:RAapprox}. The gray lines show the $95\%$ CL lower bounds achievable from the measurement of the $hVV$ couplings at current and future colliders, assuming $c_H = c_d$. {\it Right panel:} fractions for annihilation to the different SM final states. $\bar{f}f$ denotes the sum over all light quarks and leptons.}
\label{fig:fvsmchi}
\end{figure}

The ``derivative Higgs portal'' operator parametrized by $c_d$, which constitutes the only interaction between the DM and the SM contained in $\mathcal{L}_{\rm GB}$, allows the DM to annihilate to SM particles via $s$-channel Higgs exchange, and the observed DM relic density to be produced via the freeze-out mechanism. This fixes the interaction strength $c_d /f^2$ as a function of the DM mass, as shown by the blue curve in Fig.~\ref{fig:fvsmchi}, which was obtained by solving the Boltzmann equation for the $\chi$ number density using micrOMEGAs~\cite{Belanger:2018mqt}. For $m_\chi > m_h$ the relation is very simple, being approximately determined by
\begin{equation} \label{eq:RAapproxAnalytical}
1 = \frac{\Omega_{\chi + \chi^\ast}}{\Omega_{\rm DM}} \simeq \frac{\langle \sigma v_{\rm rel} \rangle_{\rm can}}{\tfrac{1}{2} \langle \sigma v_{\rm rel} \rangle}\,, \qquad \langle \sigma v_{\rm rel} \rangle \simeq \frac{c_d^2 m_\chi^2}{\pi f^4}\, 
\end{equation}
hence
\begin{equation} \label{eq:RAapprox}
\frac{f}{c_d^{1/2}} \approx 1.1\;\mathrm{TeV} \left( \frac{m_\chi}{130\;\mathrm{GeV}} \right)^{1/2},
\end{equation}
where $\langle\, \cdot \, \rangle$ denotes thermal average, $\Omega_{\rm DM}  = 0.1198\, h^{-2}$ \cite{Aghanim:2018eyx}, $\langle \sigma v_{\rm rel} \rangle_{\rm can} \approx 2 \times 10^{-26} \,\mathrm{cm}^3\, \mathrm{s}^{-1}$ is the canonical value of the thermal cross section~\cite{Steigman:2012nb}, and the dominant $\chi \chi^\ast \to WW, ZZ, hh$ channels were included in the annihilation.\footnote{The cross section for annihilation to $t\bar{t}$ scales as $\sigma_{t\bar{t}} \,v_{\rm rel} \sim N_c m_t^2 / (\pi f^4)$, as opposed to $\sigma_{WW,ZZ,hh} \,v_{\rm rel} \sim m_\chi^2 / (\pi f^4)$, therefore $t\bar{t}$ is important only for $m_\chi$ not much larger than $m_t$. See the right panel of Fig.~\ref{fig:fvsmchi}.}

Crucially, the derivative Higgs portal also leads to negligibly small cross sections for the scattering of DM with heavy nuclei: the amplitude for $q\chi \to q\chi$ scattering mediated by Higgs exchange is proportional to $|t|/f^2 \lesssim (100\;\mathrm{MeV})^2/ (1 \;\mathrm{TeV})^2 \sim 10^{-8}$, where we took $100\;\mathrm{MeV}$ as a rough estimate of the maximum momentum transfer. The expected strength of the direct detection signal is then set by the interactions contained in $\mathcal{L}_f - V_{\rm eff}$, which depend on the explicit breaking of the global symmetry.

The other important effect encapsulated in $\mathcal{L}_{\rm GB}$ is that $h$, due to its pNGB nature, has all its couplings rescaled by a universal factor with respect to their SM values: writing in unitary gauge $H = (0\,, \, \tilde{h}/\sqrt{2})^T$ we have
\begin{equation}
\tilde{h} = v + \Big( 1 - \frac{c_H}{2} \frac{v^2}{f^2} \Big)\, h\, .
\end{equation}
A robust and model-independent probe of this effect is the measurement of the $hVV$ couplings ($V=W,Z$). In Fig.~\ref{fig:fvsmchi} we compare the projected sensitivity on this observable of current and future colliders \cite{Thamm:2015zwa} with the pNGB DM parameter space, under the assumption that $c_H = c_d$.

\subsection{Explicit symmetry breaking effects}\label{subsec:Lf-Veff}
The most general effective Lagrangian coupling the pNGBs to the third generation quarks is
\begin{equation}\label{eq:Lfermions}
\mathcal{L}_f = - y_t \bar{q}_L \widetilde{H} t_R \left( 1 - \frac{c_t}{f^2} |H|^2 - \frac{c_t^\chi}{f^2} |\chi|^2 \right) -y_b \bar{q}_L H b_R \left( 1 - \frac{c_b}{f^2} |H|^2 - \frac{c_b^\chi}{f^2} |\chi |^2\right) + \text{h.c.}.
\end{equation}
The general form of the one-loop scalar potential generated by the explicit symmetry breaking is, up to quartic order in the fields,
\begin{equation}\label{eq:Veff}
V_{\rm eff} = \mu_h^2 |H|^2 + \lambda_h |H|^4 + \mu_{\rm DM}^2 |\chi |^2 +\lambda_{\rm DM} |\chi |^4 + 2 \lambda |H|^2 |\chi |^2\,.
\end{equation}
The parameters $\mu_h^2$ and $\lambda_h^2$ are fixed by requiring the observed mass and VEV for the SM-like Higgs. We only consider regions of parameters where $\langle \chi \rangle = 0$, so that $U(1)_{\rm DM}$ is not spontaneously broken and $\chi$ is stable. This imposes a mild constraint on the parameter space of the fermionic sector~(see Appendix~\ref{sec:appA} for a concrete example), whereas the gauging of $U(1)_{\rm DM}$ automatically yields $\mu^2_{\rm DM} > 0\,$.

In addition to providing the DM with a mass $m_\chi^2 = \mu_{\rm DM}^2 + \lambda v^2$, the explicit symmetry breaking can affect its phenomenology in important ways. The annihilation to SM particles is still dominated by $s$-channel Higgs exchange, but now the $\chi^\ast \chi h$ coupling has both a derivative and a non-derivative component,
\begin{equation} \label{eq:annihilationGEN}
\mathcal{M} (\chi \chi^\ast \to \mathrm{SM}) \propto \Big(c_d \frac{s}{f^2} - 2 \lambda\Big)v \simeq  \Big(c_d \frac{4m_\chi^2}{f^2} - 2 \lambda \Big)v\,.
\end{equation}
A priori, for $m_\chi > m_t$ the $\chi^\ast \chi \bar{t}t$ interaction proportional to $c_t^\chi$ can also give an important contribution to $\chi^\ast \chi \to t\bar{t}$. As we will discuss momentarily, however, in the models we consider $c_t^\chi$ is suppressed or altogether absent, hence Eq.~\eqref{eq:annihilationGEN} is a good approximation of the strength for annihilation to SM particles.

DM scattering with nuclei proceeds via $t$-channel Higgs exchange and through the contact interactions parametrized by $c_q^\chi$. The effective interactions with the SM quarks $q$ have the form
\begin{equation} \label{eq:ddGEN}
 2m_q a_q \, \bar{q}q \chi^\ast \chi\,, \qquad a_q \approx \frac{\lambda}{m_h^2} + \frac{c_q^\chi}{2f^2}\,.
\end{equation}
As already emphasized, the contribution of the derivative Higgs portal is negligible. 

Note that, for any relevant values of the parameters, the DM self-interactions mediated by $c_\chi$ and $\lambda_{\rm DM}$ are far too small to have any effects on cosmological scales.

\subsection{Origins of explicit breaking and DM scenarios}\label{subsec:scenarios}
Two irreducible sources of explicit symmetry breaking, which generate at least some of the interactions contained in Eqs.~\eqref{eq:Lfermions} and \eqref{eq:Veff}, are the gauging of the SM electroweak subgroup $SU(2)_L \times U(1)_Y \subset \mathcal{H}$ and the Yukawa couplings for the SM fermions. The SM gauging only contributes to the scalar potential and, under our assumption that the DM is a SM singlet, at one-loop level generates only $\mu_h^2$ and $\lambda_h$. In the fermion sector, Yukawas are assumed to arise via the partial compositeness mechanism~\cite{Kaplan:1991dc}: the elementary fermions couple linearly to operators of the strong sector,
\begin{equation}\label{eq:mixLagr}
\mathcal{L}^{\rm UV}_{\rm mix} \sim \lambda_q f\, \bar{q}_L \mathcal{O}_q + \lambda_t f\, \bar{t}_R \mathcal{O}_t  +  \lambda_{q^\prime} f\, \bar{q}_L \mathcal{O}_{q^{\,\prime}} + \lambda_b f\, \bar{b}_R \mathcal{O}_b + \text{h.c.},
\end{equation}
where we have ignored the flavor structure and put our focus on the masses of the third generation of quarks~\cite{Csaki:2008zd}. We have included mixings of the left-handed quark doublet with two distinct operators, as it is in general required to generate both the top and bottom Yukawa couplings. For example, in the $SO(6)/SO(5)$ and $SO(7)/SO(6)$ models the global symmetry is extended by an unbroken $U(1)_X$, hence if $t_R$ and $b_R$ are coupled to operators with different $X$ charge, two distinct embeddings of $q_L$ are needed in order to generate both $y_t$ and $y_b$. At low energies Eq.~\eqref{eq:mixLagr} leads to mass mixing between the elementary fermions and the composite resonances, and as a result the physical SM fields are linear combinations of elementary and composite degrees of freedom. Their compositeness fractions are defined schematically as $\epsilon_{L, R}^{t}\sim \lambda_{q,\,t} f / \sqrt{ m_{\ast q,\, t}^2 +  \lambda^2_{q,\,t} f^2 }$ and $\epsilon_{L, R}^{b}\sim \lambda_{q^{\prime},\,b} f / \sqrt{m_{\ast q^{\prime},\,b}^2 +  \lambda^2_{q^{\prime},\,b} f^2}$, where $m_{ \ast q,\, t, \, q^\prime, \,b}$ are the relevant masses of the resonances in the top and bottom sectors. The Yukawas have the form
\begin{equation}
\label{eq:yukawa}
y_\psi \simeq  \frac{M_{\ast \psi}}{f} \,\epsilon_L^\psi \,  \epsilon_R^\psi \,,\qquad (\psi = t, b)
\end{equation}
where $M_{\ast \psi}$ is a combination of the resonance mass parameters.

\begin{figure}[t]
\centering
\includegraphics[scale=.6]{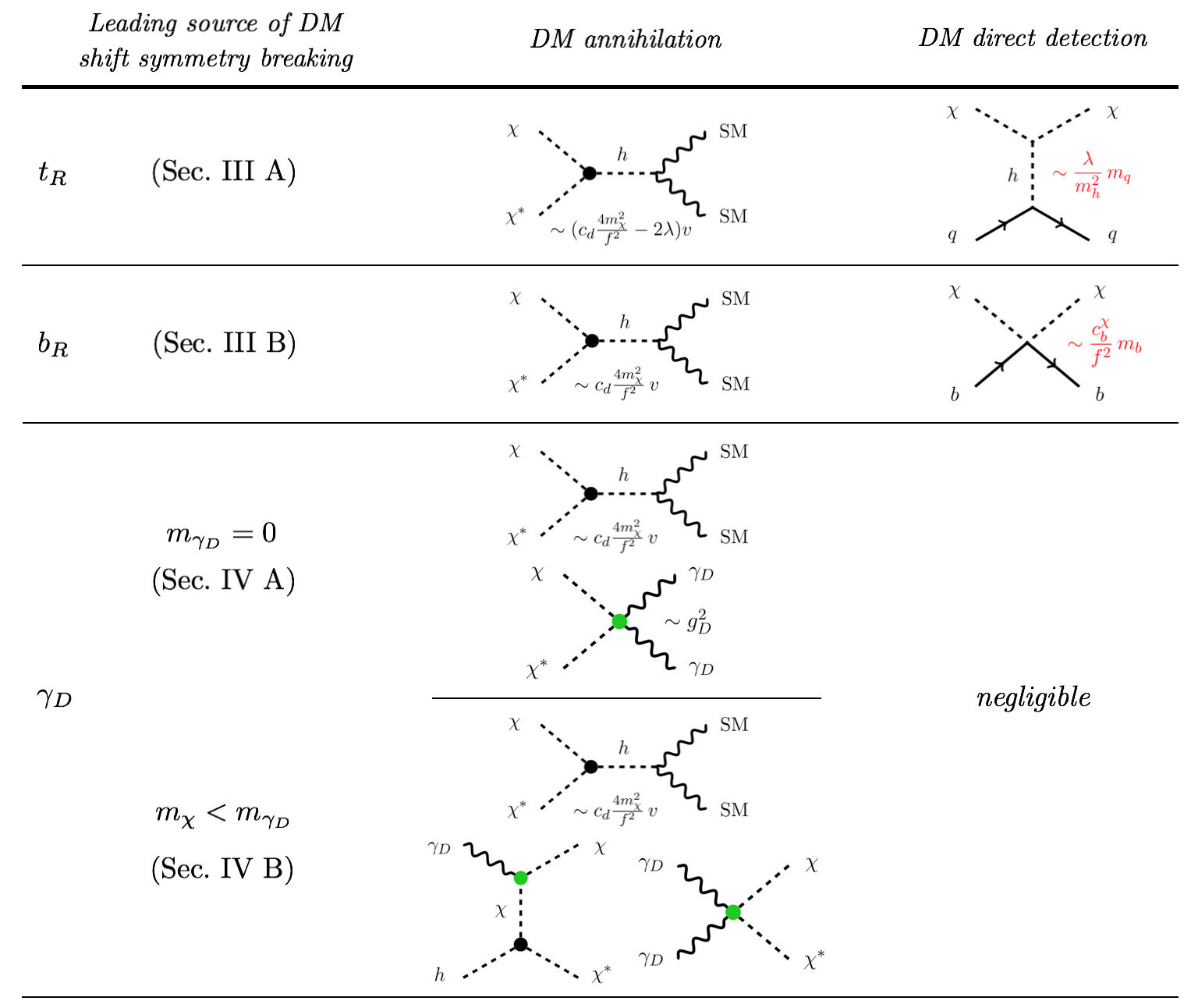}
\caption{Schematic summary of the three scenarios discussed in this paper. The EFT coefficients $c_d, c_b^\chi$ and $\lambda$ were defined in Eqs.~\eqref{eq:LGB}, \eqref{eq:Lfermions} and \eqref{eq:Veff}, respectively. In the third scenario we denote with $\gamma_D$ the dark photon associated to the gauging of $U(1)_{\rm DM}$ with coupling $g_D$, and mark the gauge interactions in green.}
\label{fig:GrandSummary}
\end{figure}

Since the elementary fermions do not fill complete $\mathcal{G}$ representations, Eq.~\eqref{eq:mixLagr} breaks explicitly at least part of the global symmetry. The Higgs shift symmetry must be broken by the couplings of both the top and bottom, in order to generate the observed values of $y_{t,b\,}$, $v$ and $m_h$. However, whether each of these couplings breaks or preserves the $\chi$ shift symmetry is a priori unknown, and all possibilities deserve close scrutiny. The three scenarios discussed in this paper are listed in Fig.~\ref{fig:GrandSummary}, along with the Feynman diagrams that dominate the annihilation and direct detection of DM in each case. In Sec.~\ref{sec:fermionbreak} we consider the scenarios where the leading breaking of the DM shift symmetry originates from the SM quarks, focusing in particular on the bottom. Then, in Sec.~\ref{sec:gaugeBreaking} we study the scenario where the fermion sector is fully symmetric, and the leading explicit breaking arises from the gauging of the $U(1)_{\rm DM}$ symmetry that stabilizes the DM. 

\section{Dark Matter shift symmetry broken by fermions}\label{sec:fermionbreak}
\noindent In this section we briefly discuss the possibility that the leading breaking of the DM shift symmetry originates from the couplings of the top quark or the bottom quark.

\subsection{Breaking of the DM shift symmetry by top quark couplings}  \label{eq:tRbreaking}
This scenario has been discussed extensively in Refs.~\cite{Frigerio:2012uc,Marzocca:2014msa,Balkin:2017aep}, and is realized e.g. for $\mathcal{O}_{q, \,t} \sim  \mathbf{7}_{2/3}$ under $SO(7)\times U(1)_X$. In this case $t_R$ interactions break the shift symmetries of $\chi$ and make the DM rather heavy, typically $m_\chi \sim 200$-$400$ GeV for $f \gtrsim \mathrm{TeV}$. At the same time the marginal Higgs portal coupling is generated with size closely related to that of the Higgs quartic, $\lambda \lesssim \lambda_h/2 \simeq 0.065$. These rough estimates imply that $ \lambda f^2 \lesssim m_\chi^2$, hence from Eq.~\eqref{eq:annihilationGEN} we read that $\lambda$ plays a subleading but non-negligible role in DM annihilation. In addition, $\lambda$ determines the DM-nucleon scattering cross section as~\cite{Balkin:2017aep}
\begin{equation} \label{eq:DDxsectionTop}
\sigma_{\rm SI}^{\chi N} \simeq \frac{f_N^2}{\pi} \frac{\lambda^2 m_N^4 }{ m_h^4 m_\chi^2} \;\approx\; 1.6 \times 10^{-46} \,\mathrm{cm}^2\, \left(\frac{\lambda}{0.02}\right)^2 \left(\frac{300\;\mathrm{GeV}}{m_\chi}\right)^2 , \qquad (t_R \;\mathrm{breaking})
\end{equation}
where $f_N \simeq 0.30$ contains the dependence on the nucleon matrix elements (since Higgs exchange dominates, all SM quarks contribute to the signal strength). The XENON1T experiment is currently probing cross sections of the size of Eq.~\eqref{eq:DDxsectionTop}, and part of the parameter space has recently been excluded by its latest results \cite{Aprile:2018dbl}. Notice that we have consistently neglected the effects of $c_t^\chi$: this is because the viable parameter space features large mixing of $t_R$ with the fermionic resonances, which strongly suppresses this coefficient~\cite{Balkin:2017aep}.

\subsection{Breaking of the DM shift symmetry by bottom quark couplings} \label{eq:bRbreaking}
A different scenario is obtained if the DM shift symmetry is fully preserved by the interactions of the top quark, but it is broken by those of the bottom. As a concrete example we take $\mathcal{O}_q \sim \mathbf{7}_{2/3}, \mathcal{O}_t \sim \mathbf{21}_{2/3}$ and $\mathcal{O}_{q^\prime, \, b} \sim \mathbf{7}_{-1/3}$ under $SO(7)\times U(1)_X$, in which case only the couplings of $b_R$ to the strong sector break the $\chi$ shift symmetries. Only the essential features of the setup are presented here, while a detailed discussion is provided in Appendix~\ref{sec:appA}. We focus on the region of parameter space where $\epsilon_L^b \sim \epsilon_R^b \sim \sqrt{y_b f / M_{\ast b}}\,$, which in turn lead to the scalings in Eq.~\eqref{eq:bottomScalings} with $g_\ast \sim M_{\ast b}/f$. As a result, the $\chi$ mass can be of $O(100)$~GeV while the portal coupling remains very suppressed. Quantitatively, we estimate
\begin{subequations}\label{eq:parametersbBreaking}
\begin{alignat}{2}
m_\chi \simeq \sqrt{\mu^2_{\rm DM}} &\,\approx\, 120\;\mathrm{GeV} \left( \frac{M_{\ast b}}{8\;\mathrm{TeV}} \right)^{3/2} \left( \frac{1\;\mathrm{TeV}}{f} \right)^{1/2},\\  \lambda &\,\approx\, 3\,\times 10^{-4}  \left( \frac{M_{\ast b}}{8\;\mathrm{TeV}} \right)^2  \left( \frac{1\;\mathrm{TeV}}{f} \right)^{2}.
\end{alignat}
\end{subequations}
The above parametrics have been confirmed by a numerical scan of the $SO(7)/SO(6)$ model whose results are reported in Appendix~\ref{sec:appA}. The important message contained in Eq.~\eqref{eq:parametersbBreaking} is that since $\lambda f^2 \ll m_\chi^2$, $\chi$ annihilation proceeds dominantly via the derivative portal, and the DM is heavy enough that the correct relic density can be reproduced for $f \sim \mathrm{TeV}$, see Fig.~\ref{fig:fvsmchi}. In addition, we have $c_b^\chi \simeq 1$ and $\lambda f^2 \ll m_h^2$ in Eq.~\eqref{eq:ddGEN}, so the scattering with nuclei is dominated by the $\chi^\ast \chi b\bar{b}$ contact interaction. The DM-nucleon scattering cross section is
\begin{align} \label{eq:DDxsectionBottom} \nonumber
\sigma_{\rm SI}^{\chi N} &\simeq \frac{\tilde{f}_N^2}{\pi} \frac{m_N^4}{ 4 f^4 m_\chi^2} \;\\&\approx\; 1.0\,\mbox{-}\, 5.6 \times 10^{-47} \,\mathrm{cm}^2\, \left(\frac{1\;\mathrm{TeV}}{f}\right)^4 \left(\frac{100\;\mathrm{GeV}}{m_\chi}\right)^2 , \quad (b_R \;\mathrm{breaking})
\end{align}
where the range of values accounts for the theory uncertainty on the couplings of the first and second generation quarks. The lower estimate corresponds to breaking of the DM shift symmetry only by the bottom quark ($c_b^\chi = 1$ and $c_q^\chi = 0$ for all $q\neq b$, case I), yielding a nucleon form factor $\tilde{f}_N \simeq 0.066$. The higher estimate corresponds to breaking by all down-type quarks ($c_{d,s,b}^\chi = 1$ and $c_{u,c,t}^\chi = 0$, case II),\footnote{This is the pattern obtained by extending the embeddings $\mathcal{O}_q \sim \mathbf{7}_{2/3}$, $\mathcal{O}_t \sim \mathbf{21}_{2/3}$ and $\mathcal{O}_{q^\prime, \, b} \sim \mathbf{7}_{-1/3}$ to all three generations.} yielding $\tilde{f}_N \simeq 0.15$. The extremely suppressed cross sections in Eq.~\eqref{eq:DDxsectionBottom} will be probed by next-generation experiments such as LZ \cite{Mount:2017qzi}, for which they constitute a very motivated target.

\begin{figure}[t]
\centering
\includegraphics[scale=.45]{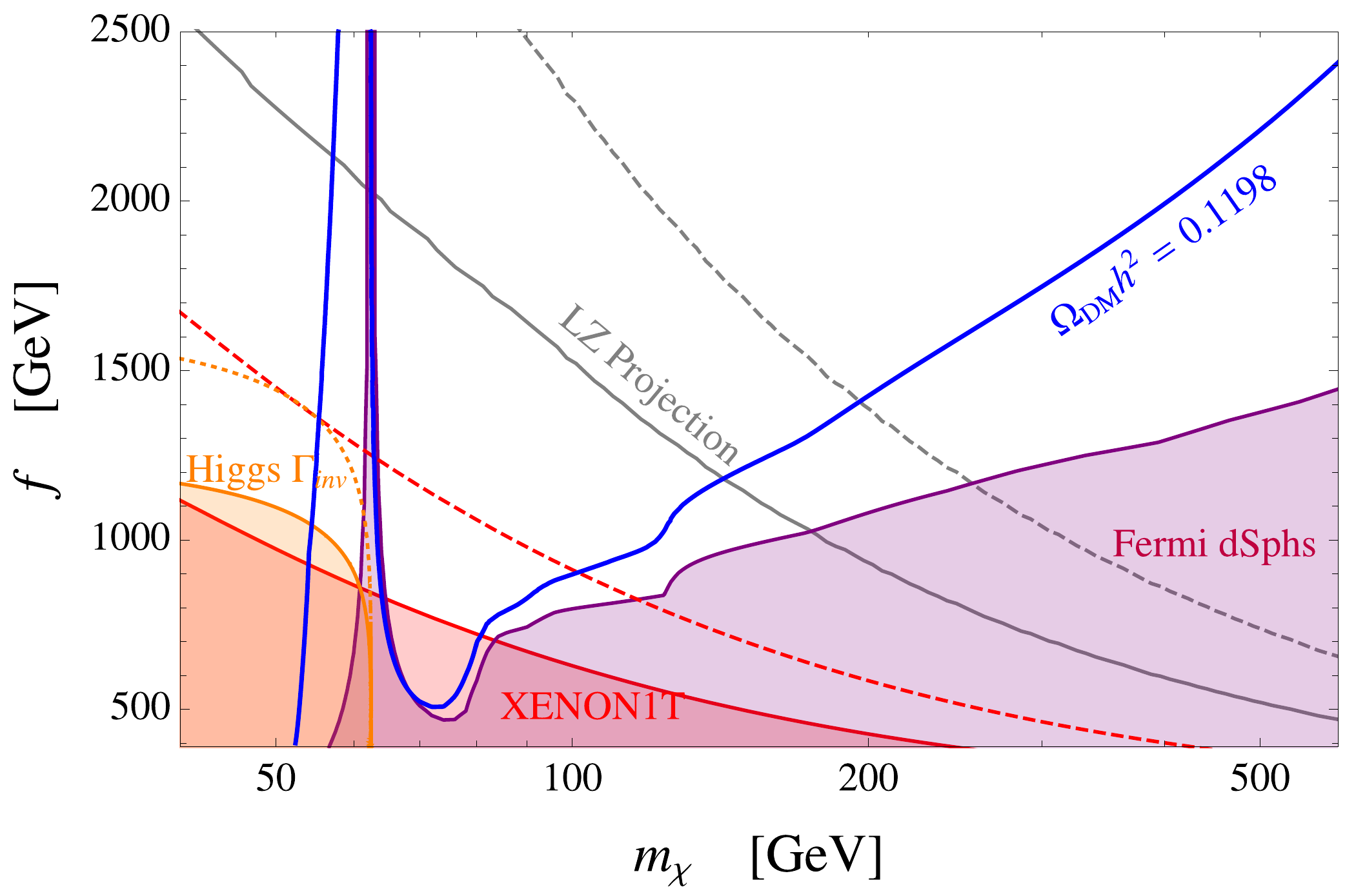}
\caption{Parameter space of the model where the bottom quark gives the leading breaking of the DM shift symmetry. The coefficients of the effective Lagrangian are set to $c_d = c_b^\chi = 1$, $c_t^\chi = \lambda = 0$. To draw the exclusions from direct and indirect detection we have assumed that all of the observed DM is composed of $\chi$ particles, irrespective of the thermal value of the $\chi$ density predicted at each $(m_\chi, f)$ point.}
\label{fig:BottomSummary}
\end{figure}

A summary of the current constraints and future reach on the $(m_\chi, f)$ parameter space is shown in Fig.~\ref{fig:BottomSummary}, where we have set $c_d = c_b^\chi = 1$, $c_t^\chi = \lambda = 0$.
Points lying on the blue curve reproduce the observed DM relic density. The red-shaded region is ruled out by current XENON1T results~\cite{Aprile:2018dbl} assuming case I for the DM-nucleon cross section, whereas the dashed red line corresponds to the exclusion for case II. The solid gray (dashed gray) lines show the expected sensitivity achieved by LZ~\cite{Mount:2017qzi} for case I (case II). The region $m_\chi < m_h/2$ is also constrained by LHC searches for invisibly-decaying Higgses. The current $95\%$ CL bound $\mathrm{BR}(h\rightarrow \chi^\ast \chi) < 0.24$ \cite{CMS:2018awd} rules out the region shaded in orange, which extends up to $f \simeq 1.2\;\mathrm{TeV}$ for very light $\chi$. The projected HL-LHC limit $\mathrm{BR}(h\to  \chi^\ast \chi) < 0.08$ \cite{ATLAShl-lhc}, corresponding to the dotted orange curve, will extend the reach to $f \simeq 1.6\;\mathrm{TeV}$. Finally, the region shaded in purple is excluded by searches for present-day DM annihilations from dwarf spheroidal galaxies (dSphs) performed at Fermi-LAT~\cite{Ackermann:2015zua}. This bound was derived by comparing the total cross section for DM annihilation in our model to the limit reported by Fermi for the $b\bar{b}$ final state, and should therefore be taken as approximate. Additional indirect detection constraints~\cite{Cuoco:2016eej,Cui:2016ppb} arise from the measurement of the antiproton spectrum by AMS-02~\cite{Aguilar:2016kjl}. These are, however, affected by systematic uncertainties whose sizes are under active debate. We have therefore chosen to quote only the more conservative bounds from dSphs. Figure~\ref{fig:BottomSummary} shows that most of the best-motivated parameter space, with $80\;\mathrm{GeV} \lesssim m_\chi \lesssim 200\;\mathrm{GeV}$ and $0.8 \;\mathrm{TeV} \lesssim f \lesssim 1.4\;\mathrm{TeV}$, is currently untested but within reach of LZ.

\section{Dark Matter shift symmetry broken by $U(1)_{\rm DM}$ gauging} \label{sec:gaugeBreaking}
\noindent It is possible to couple all the elementary quarks to the strong sector in a way that preserves the DM shift symmetry \cite{Frigerio:2012uc,Balkin:2017aep}. For example, in the $SO(7)/SO(6)$ model this is achieved with $\mathcal{O}_q \sim \mathbf{7}_{2/3}$, $\mathcal{O}_{u,d} \sim \mathbf{21}_{2/3}$ for all three generations. This setup gives $c_q^\chi = 0$ in Eq.~\eqref{eq:Lfermions} and no contribution to $\mu^2_{\rm DM}, \lambda_{\rm DM}$ and $\lambda$ in Eq.~\eqref{eq:Veff} from the fermion sector, while at the same time top loops easily produce a realistic Higgs potential. In this case, some additional explicit breaking should be responsible for generating the DM mass. If $\chi$ is a complex scalar, a natural possibility is that the explicit breaking originates from the gauging of $U(1)_{\rm DM}$. In the $SO(7)/SO(6)$ coset the generators associated with the real and imaginary parts of $\chi$ together with the $U(1)_{\rm DM}$ generator form an $SU(2)^\prime \sim \{X^{\rm Re}, X^{\rm Im}, T^{\rm DM}\}$, hence gauging $U(1)_{\rm DM}$ generates a radiative mass for $\chi$ in very similar fashion to the contribution of photon loops to the charged pion mass in the SM.

From the effective theory point of view, the effects of gauging $U(1)_{\rm DM}$ with coupling $g_D$ can be taken into account by replacing in $\mathcal{L}_{\rm GB}$ in Eq.~\eqref{eq:LGB},
\begin{equation} \label{eq:gaugingEFT}
| \partial^\mu \chi |^2 \quad \to \quad  |(\partial^\mu - i g_D A_{D}^{\mu}) \chi|^2 - \frac{1}{4} F_D^{\mu \nu} F_{D\mu \nu}   + \frac{1}{2}\, m_{\gamma_D}^2 A_{D\mu} A_{D}^{\mu}\,,
\end{equation}
where we took $\chi$ to have unit charge. Note that to be general we have included a mass term for the dark photon $\gamma_D$, which can arise via the St\"uckelberg mechanism without spontaneous breaking of $U(1)_{\rm DM}$. The one-loop DM mass and marginal portal coupling are
\begin{equation} \label{eq:potentialGauging}
m_\chi = \sqrt{ \mu_{\rm DM}^2} \simeq \sqrt{ \frac{3 \alpha_D}{2 \pi}} \,m_\rho \approx 100\;\mathrm{GeV} \left( \frac{\alpha_D}{10^{-3}} \right)^{1/2} \left( \frac{m_\rho}{5\;\mathrm{TeV}} \right), \qquad\quad  \lambda = 0\,,
\end{equation}
where $\alpha_D \equiv g_D^2 / (4\pi)$ and the loop that generates $m_\chi$ was cut off at $m_\rho$, the mass of vector resonances (in the $SO(7)/SO(6)$ model, this is the mass of the $\mathbf{15}$ multiplet of $SO(6)$). The estimate for the DM mass in Eq.~\eqref{eq:potentialGauging} is valid as long as $m_{\gamma_D} \ll m_\rho$, which we assume. Importantly, since the Higgs is uncharged under $U(1)_{\rm DM}$ the marginal portal coupling is not generated at one loop, leading from Eq.~\eqref{eq:ddGEN} to an extremely suppressed DM-nucleon cross section. We find it remarkable that such a simple model is effectively inaccessible to direct detection experiments. 

The introduction of the dark photon has significant impact on the phenomenology. It is important to stress that in Eq.~\eqref{eq:gaugingEFT} we have not included the operator $\varepsilon B_{\mu \nu} F_{D}^{\mu \nu}/2$ that mixes kinetically $U(1)_{\rm DM}$ and the SM hypercharge. The choice to set $\varepsilon = 0$ in the EFT is motivated by the $SO(7)/SO(6)$ model, where the kinetic mixing is forbidden by $C_D$, the charge conjugation of $U(1)_{\rm DM}$, which is an accidental symmetry (provided it is respected by subleading spurionic embeddings of the SM fermions, see Appendix~\ref{app:KinMixing}). In particular, in the low-energy theory $C_D$ transforms $A_{D}^\mu \to - A_{D}^\mu$ and $\chi \to - \chi^\ast$, whereas all SM fields are left unchanged. An additional, important consequence of this discrete symmetry is that the dark photon is stable if $m_{\gamma_D} < 2m_\chi$, when the $\gamma_D \to \chi \chi^\ast$ decay is kinematically forbidden. The complete discussion of kinetic mixing, as well as the details on the implementation of $C_D$ as an $O(6)$ transformation that we call $P_6$, are contained in Appendix~\ref{app:KinMixing}. 

The dark sector, composed of the DM and the dark photon, is thus characterized by the four parameters $\{m_\chi, f, \alpha_D, m_{\gamma_D}\}$. In the remainder of this section we analyze its phenomenology in detail, beginning in Sec.~\ref{subsec:masslessDP} with the simplest setup where the dark photon is massless, and later moving to the massive case in Sec.~\ref{subsec:massiveDP}.

\subsection{Phenomenology for massless dark photon} \label{subsec:masslessDP}
\begin{figure}[t]
\centering
\includegraphics[width=0.48\textwidth]{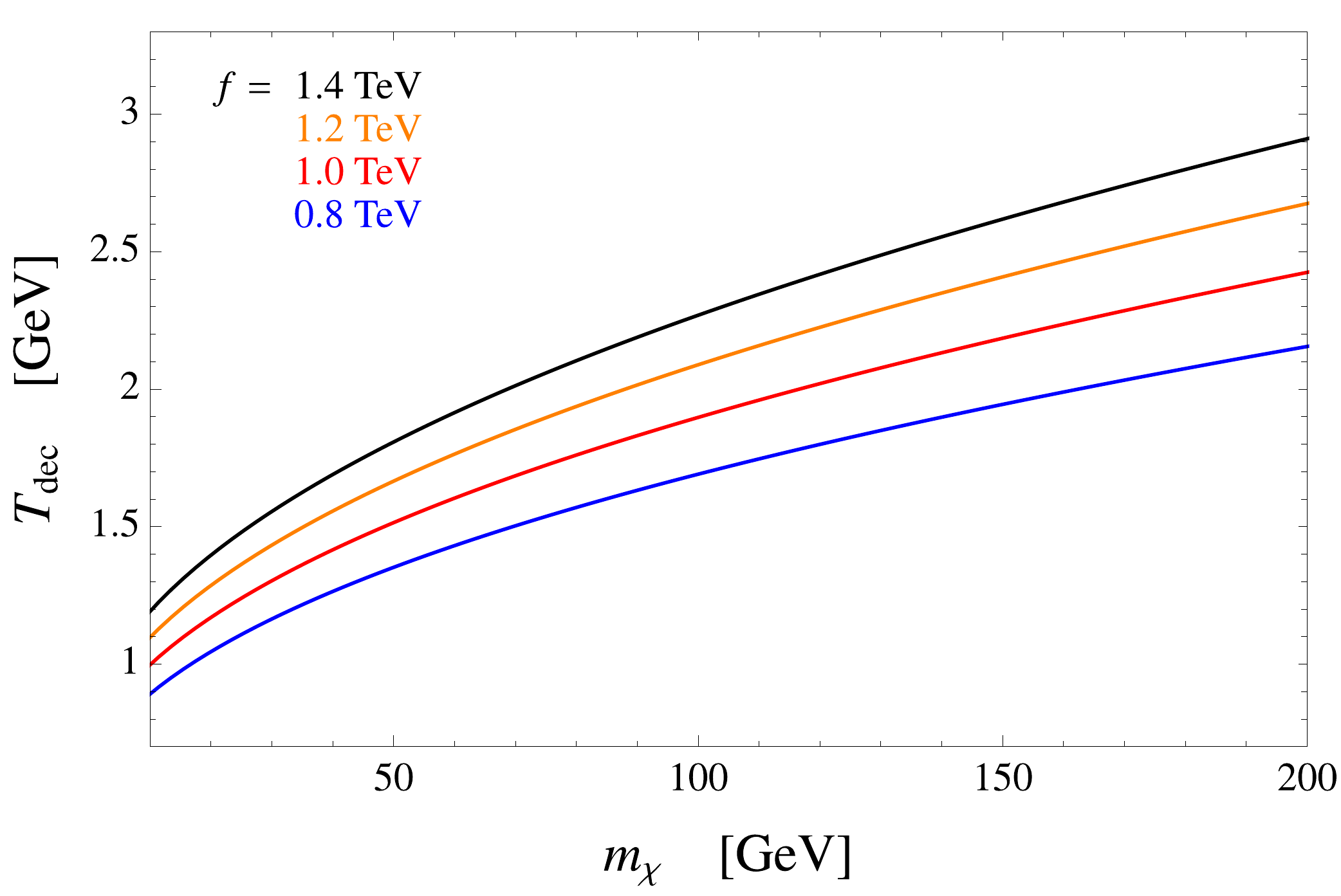}\hspace{1mm}
\includegraphics[width=0.49\textwidth]{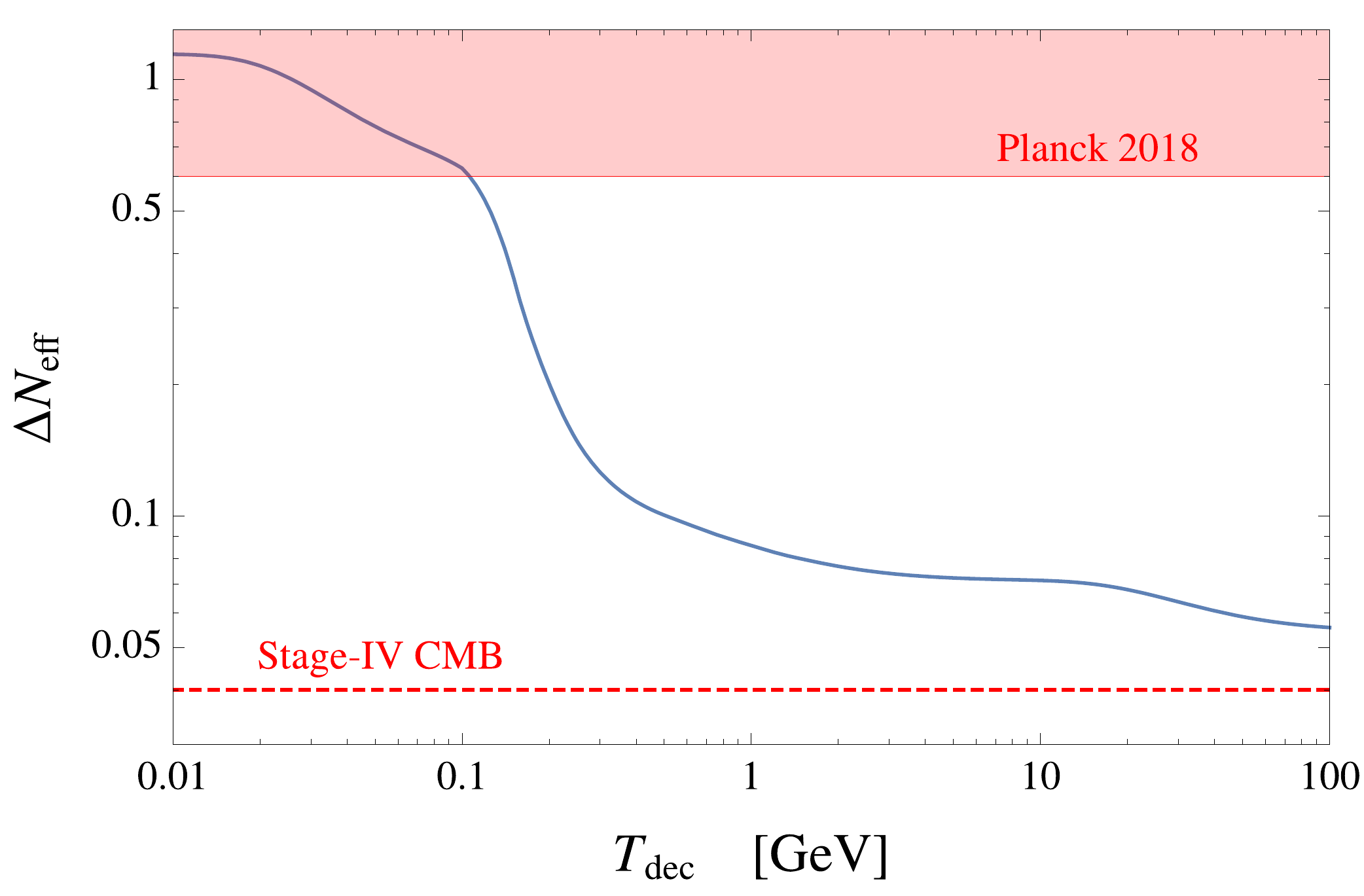}
\caption{{\it Left panel:} temperature of kinetic decoupling between the dark and visible sectors. {\it Right panel:} contribution of the dark photon to $\Delta N_{\rm eff}$ at photon decoupling, calculated from Eq.~\eqref{eq:deltaNeff}. In the evaluation of $g_{\ast s, \rm vis}(T_{\rm dec})$ we assumed $150\;\mathrm{MeV}$ as temperature of the QCD phase transition. The region shaded in red corresponds to the current CMB constraint $\Delta N_{\rm eff} \lesssim 0.6$, while the dashed red line shows the projected Stage-IV CMB bound $\Delta N_{\rm eff} \lesssim 0.04$.}
\label{fig:DarkPhoton}
\end{figure}

Setting $m_{\gamma_D} = 0$ leaves the three-dimensional parameter space $\{m_\chi, f, \alpha_D\}$. We begin the discussion with a summary of the thermal history of the model. At early times the dark sector, composed of $\chi$ and $\gamma_D$, and the visible sector are kept in kinetic equilibrium by elastic $\chi f \to \chi f$ scatterings mediated by Higgs exchange, where $f$ denotes the still-relativistic SM fermions. These processes are effective down to temperatures $T \ll m_\chi$, but eventually they become slower than the Hubble expansion rate and the dark and visible sectors decouple. The corresponding decoupling temperature $T_{\rm dec}$ is defined through \cite{Gondolo:2012vh} $H(T_{\rm dec}) = \gamma (T_{\rm dec})/2$, where $H(T) = \pi \sqrt{g_\ast (T)} \,T^2 / (3 \sqrt{10} M_{\rm Pl})$ is the Hubble parameter for a radiation-dominated Universe ($g_\ast (T)$ is the total number of relativistic degrees of freedom including both the visible and dark sectors, and $M_{\rm Pl}$ is the reduced Planck mass), whereas $\gamma (T)$ is the momentum relaxation rate, which scales as $\gamma \sim (T/ m_\chi) n_f \langle \sigma_{\chi f} v_{\rm rel} \rangle$. Using the exact expression of $\gamma (T)$ given in Ref.~\cite{Gondolo:2012vh} we calculate\footnote{For simplicity, in deriving $T_{\rm dec}$ the total number of relativistic degrees of freedom was set to the approximate constant value $g_\ast = g_{\ast, \rm vis} + g_{\rm dark} = 75.75 + 2 = 77.75$, which corresponds to $m_\tau < T < m_b$.} $T_{\rm dec}$ as a function of $m_\chi$ and $f$, finding that it is typically between $1$ and $3\;\mathrm{GeV}$ as shown in the left panel of Fig.~\ref{fig:DarkPhoton}.

The massless dark photon behaves as radiation at all temperatures. The strongest constraint on new relativistic degrees of freedom arises from Cosmic Microwave Background (CMB) measurements of the Hubble parameter, usually formulated in terms of the effective number of light neutrino species $N_{\rm eff}$. In our model the dark photon gives a contribution \cite{Ackerman:mha}
\begin{equation} \label{eq:deltaNeff}
\Delta N_{\rm eff} = N_{\rm eff} - 3.046 = \frac{8}{7}\, \frac{g_{\rm dark}(T)}{2} \left( \frac{T}{T_\nu} \right)^{4} \left( \frac{g_{\rm dark} (T_{\rm dec})}{ g_{\rm dark} (T)} \frac{g_{\ast s, \rm vis} (T)}{g_{\ast s, \rm vis} (T_{\rm dec})} \right)^{4/3} ,
\end{equation}
where $T \sim 0.3\;\mathrm{eV}$ is the photon temperature at decoupling, $N_{\rm eff} = 3.046$ is the SM prediction, $T/T_\nu = (11/4)^{1/3}$ and $g_{\ast s, \rm vis} (T) = 3.91$. To obtain Eq.~\eqref{eq:deltaNeff} we have used the fact that below $T_{\rm dec}$ the entropies of the dark and visible sectors are separately conserved. Since $\chi$ is already non-relativistic at kinetic decoupling, we have $g_{\rm dark} (T_{\rm dec}) = g_{\rm dark} (T) = 2$ and $\Delta N_{\rm eff}$ is determined by the number of SM relativistic degrees of freedom at $T_{\rm dec}$. As shown in the right panel of Fig.~\ref{fig:DarkPhoton}, as long as $T_{\rm dec} \gg 100\;\mathrm{MeV}$ the current bound $\Delta N_{\rm eff} \lesssim 0.6$ \cite{Aghanim:2018eyx} ($95\%$ CL) is easily satisfied. As we have seen, the typical decoupling temperature is $1\,$-$\,3$ GeV, corresponding to $\Delta N_{\rm eff} \approx 0.07\,$-$\,0.09$. Such values could be probed in future Stage-IV CMB measurements, which are expected to constrain $\Delta N_{\rm eff} \lesssim 0.04$ at $95\%$ CL \cite{Abazajian:2013oma}. A similar, but slightly weaker, current bound is obtained from Big-Bang nucleosynthesis \cite{Cooke:2013cba}.

In addition, the Compton scattering process $\chi \gamma_D\to \chi \gamma_D$ delays kinetic decoupling of the DM compared to the standard WIMP scenario \cite{Ackerman:mha,Feng:2009mn}, suppressing the matter power spectrum on small scales and leading to a minimum expected DM halo mass. For weak-scale DM and typical coupling $\alpha_D\sim 10^{-3}$, though, $\chi\,$-$\,\gamma_D$ kinetic decoupling takes place at temperature of $O(\mathrm{MeV})$ and the minimum halo mass is too small to be testable with current observations~\cite{Feng:2009mn}.
 
Having established that the massless dark photon does not conflict with cosmological observations, we turn to the DM phenomenology. The $\chi \chi^\ast$ pairs undergo $s$-wave annihilation both to SM particles via the derivative Higgs portal, and to $\gamma_D \gamma_D$ with amplitude mediated by the scalar QED interactions in Eq.~\eqref{eq:gaugingEFT}. The cross section for the latter is
\begin{equation} \label{eq:annihilationDarkPhoton}
\langle \sigma_{\gamma_D \gamma_D} v_{\rm rel} \rangle = \frac{2\pi \alpha_D^2}{m_\chi^2}\,,
\end{equation}
where we took the leading term in the velocity expansion. Notice that the ``mixed'' dark-visible annihilation $\chi \chi^\ast \to \gamma_D h$ is instead $p$-wave suppressed: the amplitude vanishes at threshold, because spin cannot be conserved for $m_{\gamma_D} = 0$.\footnote{The $p$-wave suppression applies also for $m_{\gamma_D} \neq 0$, since the longitudinal polarization does not contribute to the amplitude due to $U(1)_{\rm DM}$ invariance.\vspace{1.5mm}} Therefore this process has only a very small impact on the freeze-out. The requirement to obtain the observed relic density yields a two-dimensional manifold in the parameter space, whose features are best understood by considering slices with fixed $f$. 

As discussed in Sec.~\ref{subsec:LGB}, there exists then only one value of the DM mass which gives the correct relic density by annihilation only through the derivative Higgs portal: for example, for $f = 1\,(1.4)\,\mathrm{TeV}$ this is $m_\chi^{(f)} \approx 122\,(194)\,\mathrm{GeV} $. For $m_\chi > m_\chi^{(f)}$ the derivative portal coupling strength $\sim m_\chi^2 / f^2$ is too large, yielding DM underdensity for any value of $\alpha_D$. Conversely, for $m_\chi <m_\chi^{(f)}$ the $\chi \chi^\ast \to \gamma_D \gamma_D$ annihilation compensates for the reduced derivative portal for an appropriate value of $\alpha_D$. Comparing Eqs.~\eqref{eq:RAapproxAnalytical} and \eqref{eq:annihilationDarkPhoton}, the two annihilation channels have equal strength when $\alpha_D^2 \sim m_\chi^4 / (2\pi^2 f^4)$, which since $m_\chi /f \sim 1/10$ corresponds to $\alpha_D \sim 2 \times  10^{-3}$. For very light DM, $m_\chi \ll m_h/2$, only annihilation to dark photons is relevant and the coupling is fixed to $\alpha_D \approx 7 \times 10^{-4}\, (m_\chi / 30\;\mathrm{GeV})$ by the analog of Eq.~\eqref{eq:RAapproxAnalytical}. 
\begin{figure}
  \begin{minipage}[b]{0.49\textwidth}
    \includegraphics[width=\textwidth]{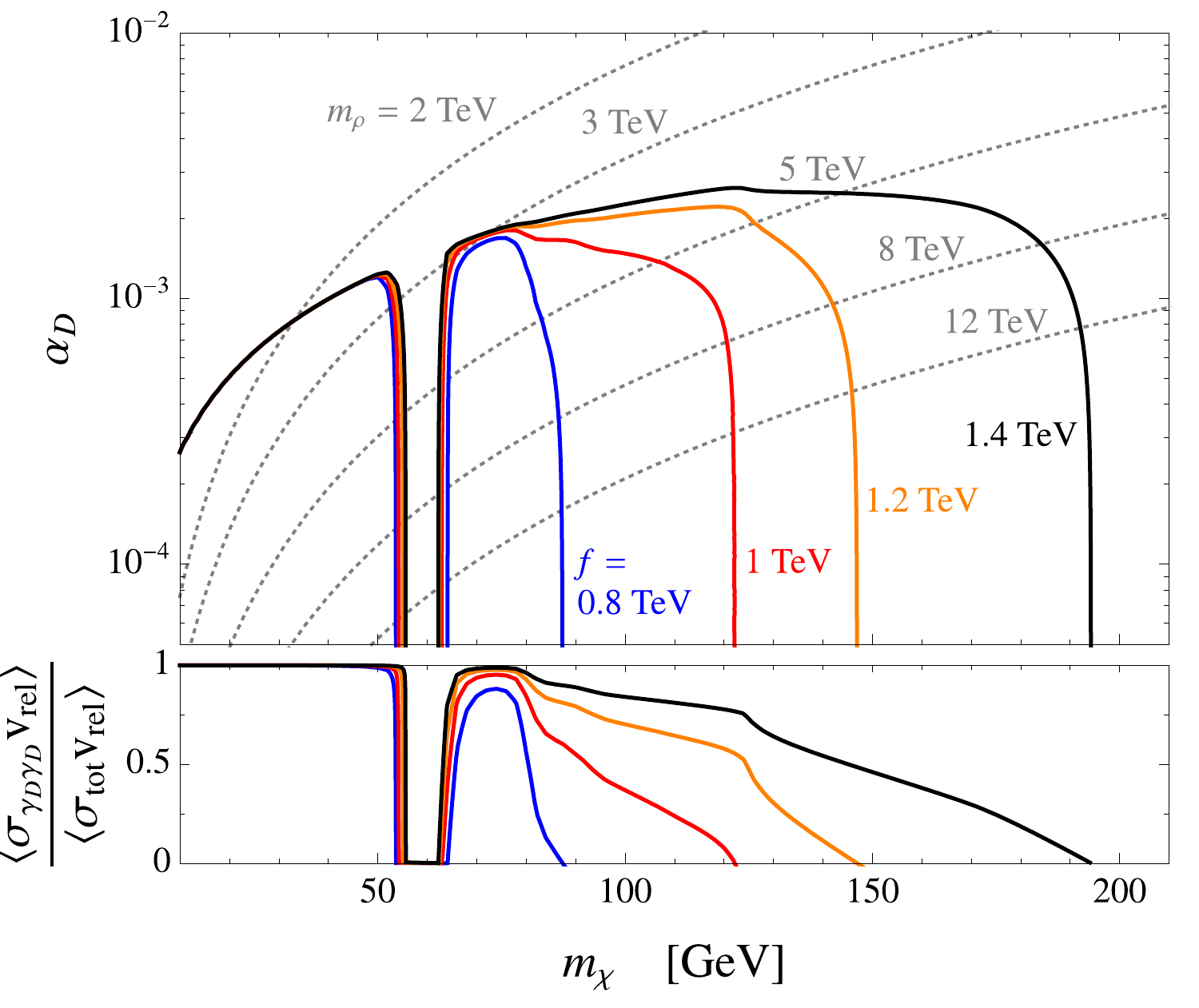}
    \label{fig:1}
  \end{minipage}
   \begin{minipage}[b]{0.49\textwidth}
    \includegraphics[width=\textwidth]{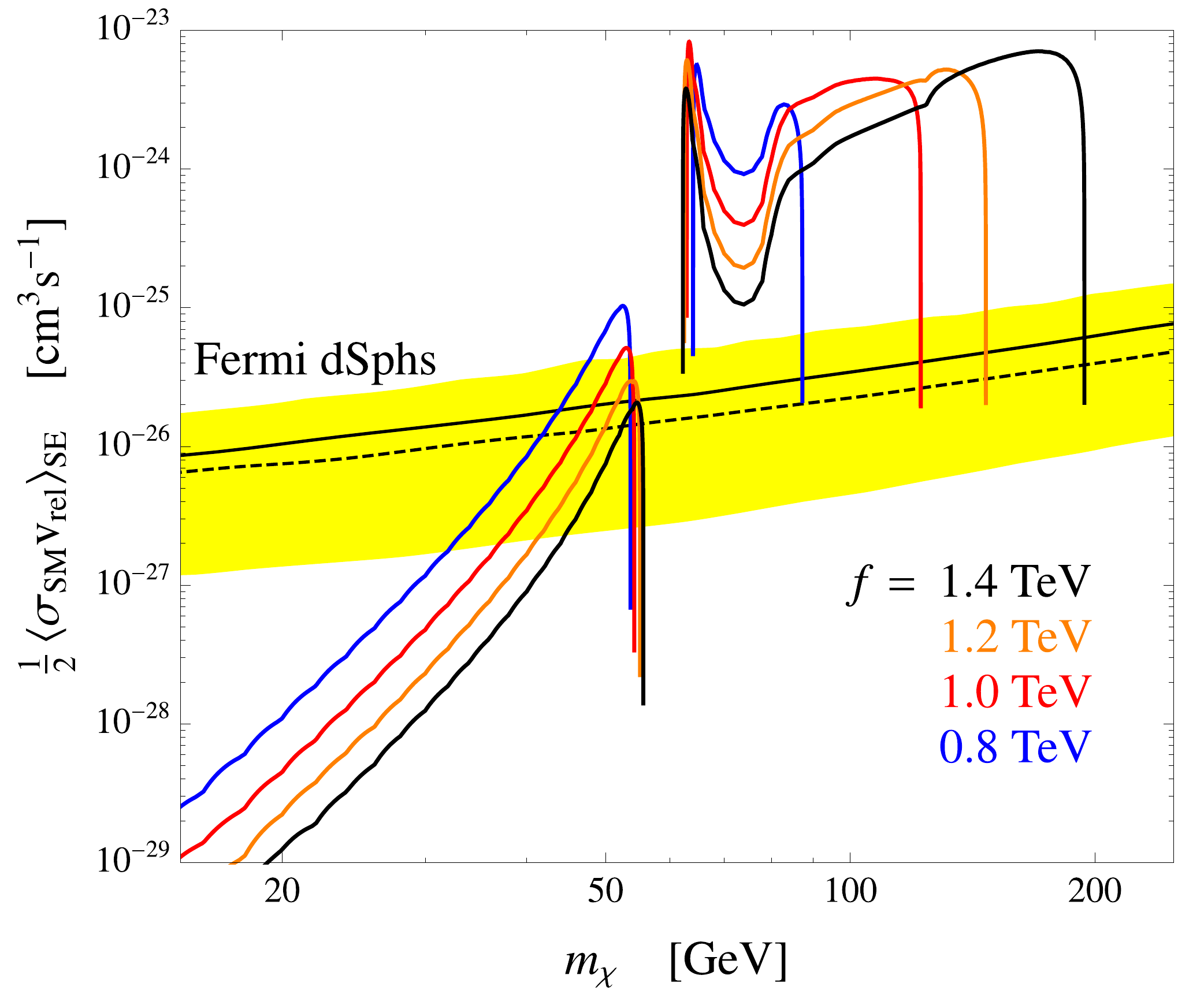}
    \vspace{+0.0cm}
    \label{fig:2}
  \end{minipage}
  \vspace{-1cm}
  \caption{{\it Left panel:} contours of observed DM relic density for representative values of $f$. The inset shows the fraction of annihilations to dark photons. Contours of constant vector resonance mass $m_\rho$ are also shown, as dashed grey lines. {\it Right panel:} the colored curves show $\langle \sigma_{\rm SM} v_{\rm rel} \rangle_{\rm SE}$, the present-day annihilation cross section to SM particles including Sommerfeld enhancement, calculated along the relic density contours shown in the left panel. The black line is the observed $95\%$ CL upper limit from the dSphs analysis in Ref.~\cite{Fermi-LAT:2016uux}. The yellow band corresponds to $95\%$ uncertainty on the expected limit in the same analysis. We also show, as dashed black line, the observed limit from the analysis of a smaller dSphs sample \cite{Ackermann:2015zua}. The quoted experimental limits were obtained assuming DM annihilation to $b\bar{b}$.}
  \label{fig:RD+ID}
\end{figure}
These features are illustrated by the left panel of Fig.~\ref{fig:RD+ID}, where contours of the observed relic abundance in the $(m_\chi, \alpha_D)$ plane are shown. Notice that in the window $55\;\mathrm{GeV} \lesssim m_\chi \lesssim 62.5\;\mathrm{GeV}$ the DM is always underdense, because the annihilation to SM particles is too strongly enhanced by the Higgs resonance. To help identify the most plausible parameter space we also show contours of constant vector resonance mass $m_\rho$, as obtained from the one-loop expression of the $\chi$ mass in Eq.~\eqref{eq:potentialGauging}.\footnote{Precisely, we employed Eq.~\eqref{eq:gaugeDMmass} with $f_\rho = f$.} We expect $1 \lesssim m_\rho /f \lesssim 4\pi$, although stronger lower bounds can arise from electroweak precision tests and from direct searches for the $\rho$ particles at colliders.

The massless dark photon mediates a long-range force between DM particles, which leads to the non-perturbative Sommerfeld enhancement (SE)~\cite{Sommerfeld:1931} of the annihilation cross section. For $s$-wave annihilation the cross section times relative velocity including SE is
\begin{equation} \label{eq:SECoulomb}
(\sigma v_{\rm rel})_{\rm SE} = (\sigma v_{\rm rel})_0 \,S(\alpha_D / v_{\rm rel}), \qquad\quad S(\zeta) = \frac{2\pi \zeta }{1 - e^{-2 \pi \zeta}} \,,
\end{equation}
where $ (\sigma v_{\rm rel})_0$ is the perturbative result, e.g. $ (\sigma v_{\rm rel})_0 = 2\pi \alpha_D^2 / m_\chi^2$ for $\chi \chi^\ast \to \gamma_D \gamma_D$. The SE is important when the ratio $\alpha_D /v_{\rm rel}$ is not too small, and scales as $S \simeq 2\pi \alpha_D/ v_{\rm rel}$ for $\alpha_D/v_{\rm rel} \gtrsim 1/2$. Assuming a Maxwell-Boltzmann distribution for the DM velocity, the thermally averaged cross section times relative velocity including SE can be written in the approximate form~\cite{Feng:2010zp}
\begin{equation} \label{eq:SEthermalAverage}
\langle \sigma v_{\rm rel} \rangle_{\rm SE} =  (\sigma v_{\rm rel})_0\, \overline{S}_{\rm ann}\,, \qquad \overline{S}_{\rm ann} = \sqrt{\frac{2}{\pi}} \frac{1}{v_0^3 N} \int_{0}^{v_{\rm max}} dv_{\rm rel} \, S(\alpha_D / v_{\rm rel})  v_{\rm rel}^2 \,e^{- \frac{v_{\rm rel}^2}{2 v_0^2}}
\end{equation}
where $v_0$ is the most probable velocity. The maximal relative velocity $v_{\rm max}$ and the normalization constant $N$ depend on whether we consider early-Universe annihilation around the time of freeze-out, in which case $v_{\rm max} = \infty$ and $N = 1$, or present-day annihilation in a galaxy halo, where $v_{\rm max} = 2 \,v_{\rm esc}$ with $v_{\rm esc}$ the escape velocity, and $N = \mathrm{erf} ( z/ \sqrt{2} ) - \sqrt{ 2/ \pi }\,z\, e^{- z^2/2}\,$, $z \equiv v_{\rm max} / v_0$. We have checked that Eq.~\eqref{eq:SEthermalAverage} agrees within a few percent with the full numerical treatment. At DM freeze-out the typical DM speed is $v_0 = \sqrt{2 / x_{\rm fo}} \sim 0.3$ since $x_{\rm fo} \equiv m_\chi / T_{\rm fo} \sim 25$, so for the typical coupling $\alpha_D \sim 10^{-3}$ the SE enhancement is negligible. Today, however, DM particles are much slower, with typical relative velocities of $10^{-3}$ in the Milky Way (MW), and $\lesssim 10^{-4}$ in dwarf galaxies. For the MW we take $v_0 = 220\,\mathrm{km/s}$ and $v_{\rm esc} = 533\,\mathrm{km/s}$ \cite{Piffl:2013mla}, obtaining a typical SE of $\overline{S}_{\rm ann} \approx 6.9$ for $\alpha_D = 10^{-3}$. For a dwarf galaxy with representative parameters $v_0 = 10\,\mathrm{km/s}$ and $v_{\rm esc} = 15\,\mathrm{km/s}$ \cite{Cirelli:2016rnw} we find $\overline{S}_{\rm ann} \approx 150$, again for $\alpha_D = 10^{-3}$. If the DM has a sizeable annihilation to SM particles, these large enhancements lead to conflict with bounds from indirect detection of DM. 

The strongest constraint comes from the non-observation by the Fermi-LAT \cite{Ackermann:2015zua,Fermi-LAT:2016uux} of excess gamma ray emission from dSphs, which are the most DM-dominated galaxies known. For $m_\chi \sim 100\;\mathrm{GeV}$ the current exclusion on $\langle \sigma v_{\rm rel} \rangle$ is about the thermal relic value. In the right panel of Fig.~\ref{fig:RD+ID} we show the total cross section for $\chi$ annihilation to SM particles, including the SE, calculated along contours in the $\{m_\chi, f, \alpha_D\}$ parameter space where the observed relic density is reproduced. Due to the large SE the region $m_\chi > m_h/2$, where an $O(1)$ fraction of DM annihilations produce SM particles, is ruled out by dSphs analyses. Notice that the experimental limits shown in Fig.~\ref{fig:RD+ID} were obtained assuming DM annihilates to $b\bar{b}$ only, whereas our $\chi$ annihilates to a combination of SM final states (see the right panel of Fig.~\ref{fig:fvsmchi}), but the uncertainty due to this approximation is mild and cannot change the conclusion that the region $m_\chi > m_h/2$ is excluded. Furthermore, in our analysis we have neglected the effects of bound state formation, which has the same parametric dependence on $\alpha_D / v_{\rm rel}$ as the SE and is expected to further enhance the signal from dSphs by an $O(1)$ factor (see Ref.~\cite{Cirelli:2016rnw} for a comprehensive analysis). On the other hand, bound state formation has negligible impact on freeze-out for the relatively light DM we consider in this work, $m_\chi \sim 100\;\mathrm{GeV}$ \cite{vonHarling:2014kha}.

Additional, important constraints on the DM self-interaction mediated by the dark photon arise from observations of DM halos. The strongest such bounds come from the triaxial structure of galaxy halos, in particular from the well-measured nonzero ellipticity of the halo of NGC720 \cite{Buote:2002wd}. This disfavors strong self interactions, which would have reduced the anisotropy in the DM velocity distribution via the cumulative effect of many soft scatterings \cite{Feng:2009mn}. In the nonrelativistic limit the scattering of two DM particles is dominated by dark photon exchange. The differential cross section in the center of mass frame is
\begin{equation} \label{eq:SIDMxsec}
\frac{d \sigma}{d \Omega} \simeq \frac{\alpha_D^2}{4m_\chi^2 v_{\rm cm}^4 (1 - \cos \theta_{\rm cm})^2}
\end{equation}
where we only retained the leading singular behavior at small $\theta_{\rm cm}$, which is the same for same-charge $\chi \chi \to \chi \chi$ and opposite-charge $\chi \chi^\ast \to \chi \chi^\ast$ scattering. Notice the very strong velocity dependence $\propto v_{\rm cm}^{-4}$, which implies that constraints from galaxies are much stronger than those from clusters. The authors of Ref.~\cite{Feng:2009mn} obtained a constraint by requiring that the relaxation time to obtain an isotropic DM velocity distribution be longer than the age of the Universe, 
\begin{equation}
\tau_{\rm iso} \equiv  \langle E_k \rangle / \langle \dot{E}_k \rangle  = \mathcal{N} m_{\chi}^3 v_0^3 (\log \Lambda)^{-1} / ( \sqrt{\pi} \alpha_D^2 \rho_\chi )  > 10^{10}\,\mathrm{years}
\end{equation}
where $E_k = m_\chi v^2/2$, $\dot{E}_k$ is the rate of energy transfer proportional to $d\sigma/d\Omega$, $\mathcal{N}$ is an $O(1)$ numerical factor, $v_0$ is the velocity dispersion (very roughly $250\,\mathrm{km/s}$ in NGC720), $\rho_\chi = m_\chi n_\chi$ is the $\chi$ energy density and the ``Coulomb logarithm'' $\log \Lambda$ originates from cutting off the infrared divergence arising from Eq.~\eqref{eq:SIDMxsec}. The ellipticity bound was recently reconsidered by the authors of Ref.~\cite{Agrawal:2016quu}, who found it to be significantly relaxed compared to the original calculation of Ref.~\cite{Feng:2009mn}. We do not review their thorough analysis here, but simply quote the result
\begin{equation}\label{eq:ellipticity}
\alpha_D < 2.4 \times 10^{-3} \left( \frac{m_\chi}{100\;\mathrm{GeV}} \right)^{3/2}. \qquad (\mathrm{ellipticity})
\end{equation}
Although Ref.~\cite{Agrawal:2016quu} considered Dirac fermion DM, their ellipticity bound directly applies to our model, because the leading term of the self-scattering cross section in Eq.~\eqref{eq:SIDMxsec} is the same for fermions and scalars.\footnote{Notice that Fig.~4 in Ref.~\cite{Agrawal:2016quu} was drawn requiring $\Omega_{X} = 0.265$ for the DM density, instead of the correct $2\,\Omega_X = 0.265$. As a result, for $m_X < 200\;\mathrm{GeV}$ (where the SE is negligible) their relic density contour should be multiplied by $\sqrt{2}$. We thank P.~Agrawal for clarifications about this point.} Furthermore, there exist several reasons \cite{Agrawal:2016quu} to take even the bound in Eq.~\eqref{eq:ellipticity} with some caution, including the fact that it relies on a single galaxy, and that the measured ellipticity is sensitive to unobservable initial conditions (for example, a galaxy that recently experienced a merger may show a sizeable ellipticity even in the presence of strong DM self-interactions). Therefore we also quote the next most stringent constraint, obtained by requiring that the MW satellite dSphs have not evaporated until the present day as they traveled through the Galactic DM halo \cite{Kahlhoefer:2013dca}. This yields
\begin{equation} \label{eq:dSphsEvap}
\alpha_D < 5 \times 10^{-3} \left( \frac{m_\chi}{100\;\mathrm{GeV}} \right)^{3/2}, \qquad (\mathrm{dwarf\;survival})
\end{equation}
which stands on a somewhat more robust footing than ellipticity, but is not free from caveats either~\cite{Agrawal:2016quu}.

\begin{figure}[t]
\centering
\includegraphics[width=0.495\textwidth]{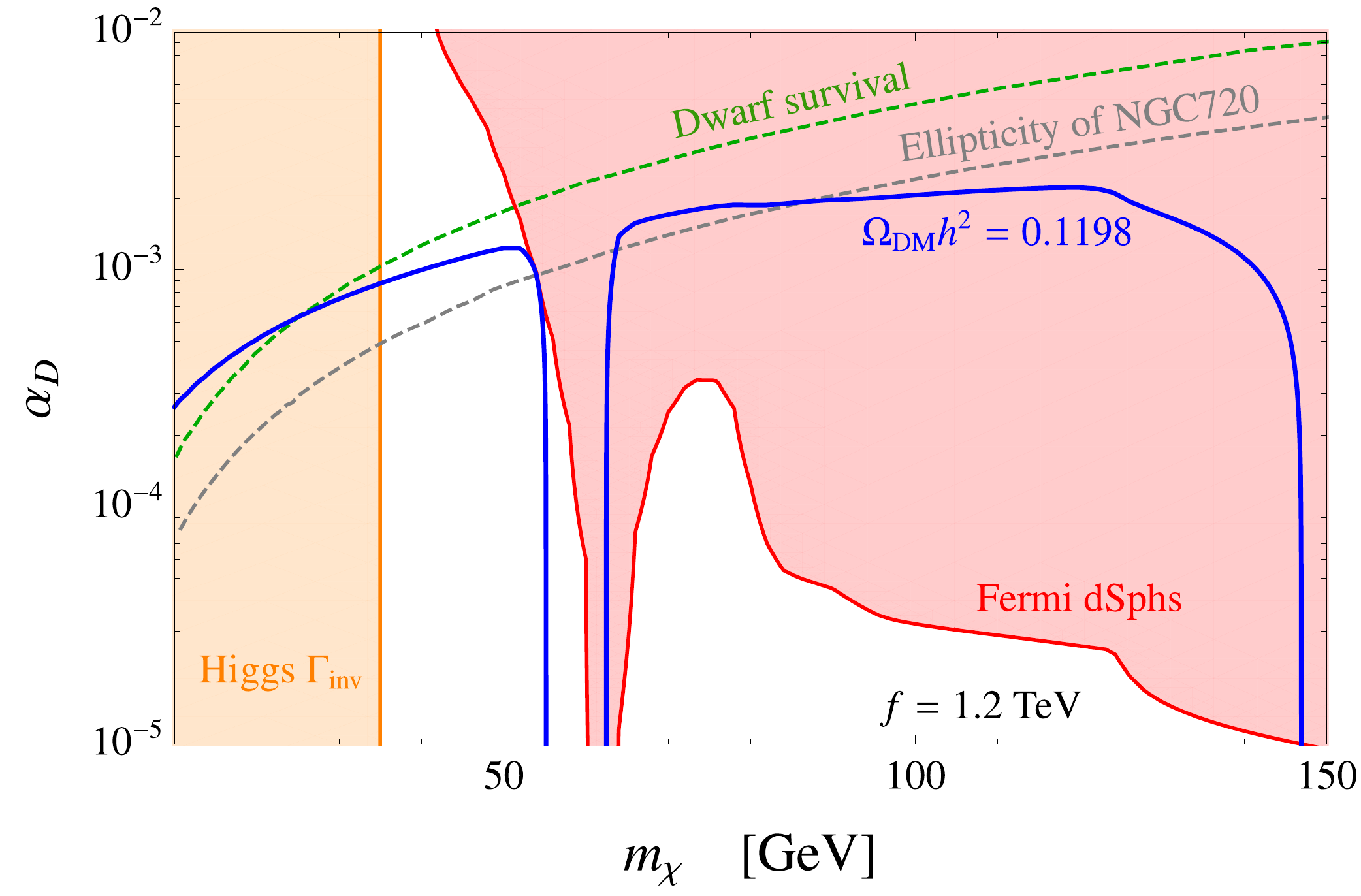}
\includegraphics[width=0.495\textwidth]{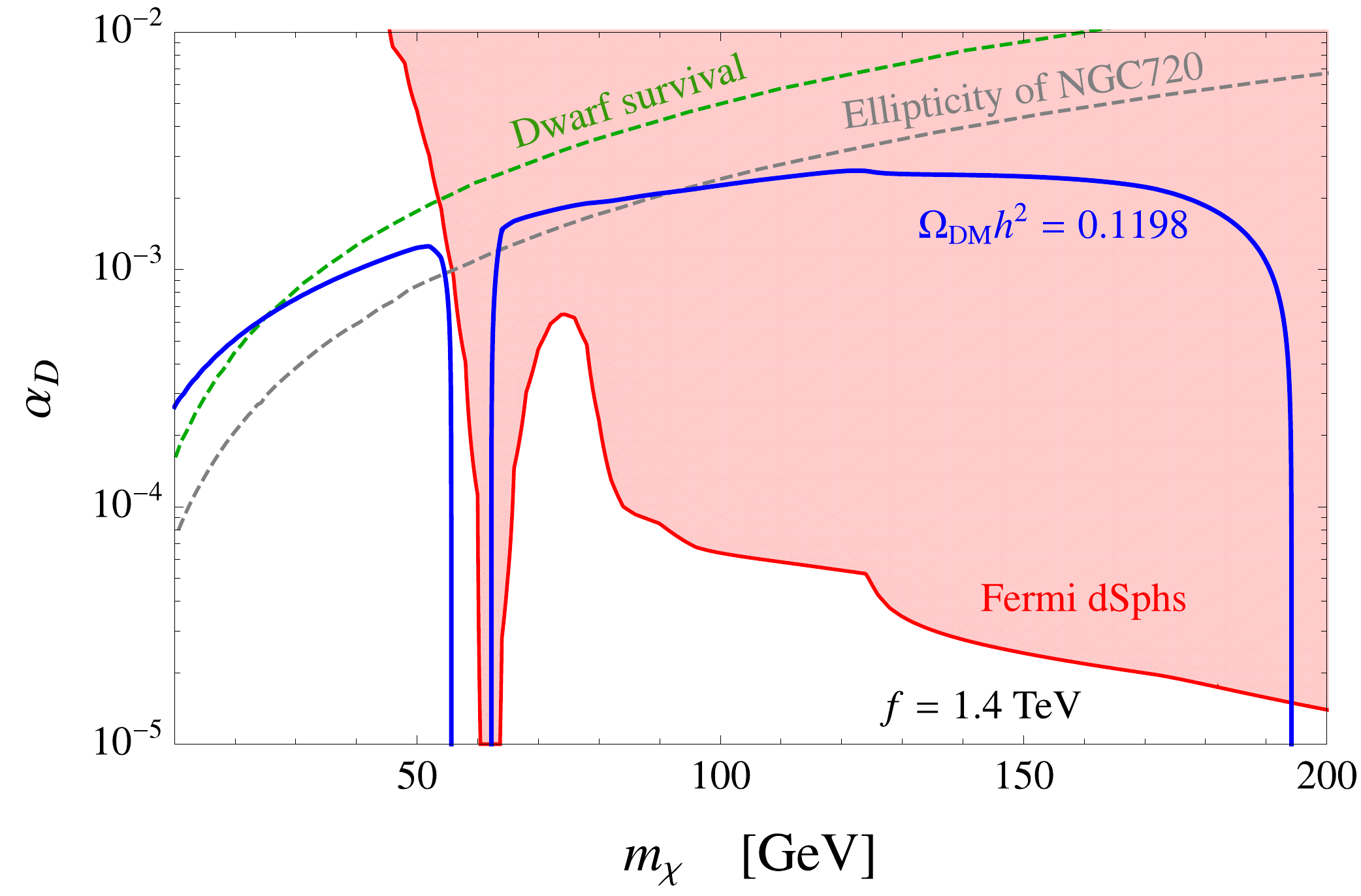}
\caption{Parameter space of the model where the gauging of $U(1)_{\rm DM}$ gives the leading breaking of the DM shift symmetry, for $f = 1.2\;\mathrm{TeV}$ (left panel) and $f = 1.4\;\mathrm{TeV}$ (right panel). The coefficients of the effective Lagrangian are set to $c_d = 1$, $c_t^\chi = c_b^\chi  = \lambda = 0$, $m_{\gamma_D} = 0$. The exclusions from Fermi dwarfs were drawn assuming that all of the observed DM is composed of $\chi$ particles, irrespective of the thermal value of the $\chi$ density predicted at each point in parameter space.}
\label{fig:masslessDPsummary}
\end{figure}
A summary of all constraints on our parameter space is shown in Fig.~\ref{fig:masslessDPsummary}, for the choices $f = 1.2$ and $1.4\;\mathrm{TeV}$. While the region $m_\chi > m_h/2$ is ruled out by gamma ray observations from dSphs, for $m_\chi < m_h/2$ the strongest bounds arise from ellipticity and dwarf evaporation. In light of the previous discussion, however, we do not interpret these as strict exclusions, but rather note that they constitute an important class of probes of our setup, which may in the near future provide important evidence in favor of, or against, DM self-interactions mediated by a massless dark photon. Such self-interactions could also have interesting implications~\cite{Agrawal:2016quu} for the small-scale issues of the collisionless cold DM paradigm~\cite{Tulin:2017ara}. A complementary test of the light DM mass region is the search for invisible $h\to \chi^\ast \chi$ decays at the LHC,\footnote{The Higgs can also decay to $\gamma_D \gamma_D$ via a $\chi$ loop. The decay width for $m_{\gamma_D} = 0$ is $\Gamma(h\to \gamma_D \gamma_D) = m_h^3 \alpha_D^2 c_d^2 v^2 | F\big(\tfrac{m_h^2}{4m_\chi^2}\big) |^2 / (64\pi^3 f^4)$, where $F(\tau)$ is given in Eq.~\eqref{eq:hgaDgaD}. Numerically, for $m_\chi < m_h/2$ this is negligible compared to $\Gamma(h \to \chi^\ast \chi)$, while for $m_\chi > m_h/2$ it is too small to be observable: e.g. for $m_\chi = 100\;\mathrm{GeV}$ and $f = 1\;\mathrm{TeV}$ we have $\Gamma(h\to \gamma_D \gamma_D) \sim 10^{-12} \;\mathrm{GeV}$.} which will be sensitive to $f\lesssim 1.6\;\mathrm{TeV}$ by the end of the high-luminosity phase (see Fig.~\ref{fig:BottomSummary}).

\subsection{Phenomenology for massive dark photon} \label{subsec:massiveDP}
\enlargethispage{-5mm}We regard the mass of the dark photon as a free parameter of our model. Having extensively discussed the simplest possibility $m_{\gamma_D} = 0$ in Sec.~\ref{subsec:masslessDP}, we turn here to the study of the massive case. The physics is qualitatively different if $m_{\gamma_D} < m_\chi$ or $m_{\chi} < m_{\gamma_D}$, so we analyze these two regions separately. Our main findings are that (1) the region $m_{\gamma_D} < m_\chi$ is ruled out, unless $\gamma_D$ is so light that it still behaves as radiation today, and (2) for $m_\chi \lesssim m_{\gamma_D} < 2m_\chi$ we obtain a two-component DM setup with novel properties. Table~\ref{Tab:summaryDarkPhoton} summarizes the mileposts in the $m_{\gamma_D}$ parameter space.
\begin{table}
   \begin{center}
   \setlength{\doublerulesep}{0.04pt}
   \begin{tabular}{ccc}
    \hline
   && \\[-0.7cm]
    \multirow{2}{*}{$m_{\gamma_D} < 6 \times 10^{-4} \;\mathrm{eV}$} & \multirow{2}{*}{$\;\; \checkmark /\, \text{X} \;\;$} & $\gamma_D$ is dark radiation today, \\ 
     & &$\;$ strong constraints from SE of $\chi \chi^\ast \to \mathrm{SM} \;$  \\      
   && \\[-0.7cm]  
    \hline
     \multirow{2}{*}{$\;6 \times 10^{-4} \;\mathrm{eV} < m_{\gamma_D} \lesssim 3m_\chi / 25\;$} & \multirow{2}{*}{$\;\; \text{X} \;\;$} & $\gamma_D$ is relativistic at freeze-out, \\ 
     & & ruled out by warm DM bounds/overabundant \\     
   && \\[-0.7cm]  
    \hline
 $ 3m_\chi / 25 < m_{\gamma_D} < m_\chi$ & $\;\; \text{X} \;\;$ & $\;\gamma_D$ is non-relativistic at freeze-out, overabundant $\,$ \\      
   && \\[-0.7cm]  
 \hline\hline
 $ m_\chi \lesssim m_{\gamma_D} < 2 m_\chi$ & $\;\; \checkmark \;\;$ & both $\gamma_D$ and $\chi$ are cold DM  \\ 
   && \\[-0.7cm]  
 \hline
$2 m_\chi < m_{\gamma_D} $ & $\;\; \checkmark \;\;$ & $\gamma_D$ is unstable  \\ 
\hline
   \end{tabular}   
%   }
   \end{center}
   \caption{Overview of the different regions in the dark photon mass space. The second column indicates whether each region satisfies ($\checkmark$) or conflicts with ($\text{X}$) experimental constraints, while the third column summarizes the key features.}
   \label{Tab:summaryDarkPhoton}
\end{table}

\subsubsection{Light dark photon: $m_{\gamma_D} < m_\chi$}
If $m_{\gamma_D} < m_\chi$, the dark photon abundance freezes out almost simultaneously with the $\chi$ abundance. Assuming $\gamma_D$ is still relativistic at freeze-out, i.e. $m_{\gamma_D} \lesssim 3 T_{\rm fo}^\chi \approx  3 m_\chi / 25$, the ratio of its number density to the SM entropy density $s_{\rm SM} = (2 \pi^2 /45) g_{\ast s, \rm vis} T^3$ is $r_{\gamma_D} = n_{\gamma_D} / s_{\rm SM} = 45\, \zeta (3) g_{\gamma_D} / ( 2\pi^4 g_{\ast s, \rm vis}) \approx 0.01$, where we assumed that the dark and visible sectors are still in kinetic equilibrium at freeze-out, and took $g_{\gamma_D} = 3$, $g_{\ast s, \rm vis} \sim 80$. Since after freeze-out there are no $\gamma_D$-number-changing interactions in equilibrium (the scattering $\gamma_D \chi \to (h^\ast \to f\bar{f}\,) \chi$ is extremely suppressed), $r_{\gamma_D}$ is conserved.\footnote{Before kinetic decoupling of the dark and visible sectors only $n_{\gamma_D}/s_{\rm tot}$ is conserved, where $s_{\rm tot}$ is the total entropy, but $s_{\rm tot} \approx s_{\rm SM}$ since $g_{\gamma_D} \ll g_{\ast s, \rm vis}$.} As the Universe cools the dark photon becomes non relativistic, its energy density being $\Omega_{\gamma_D} = m_{\gamma_D} r_{\gamma_D} s_{\rm SM}$. Requiring that today this does not exceed the observed DM density yields
\begin{equation} \label{eq:overclosureBound}
\Omega_{\gamma_D} < \Omega_{\rm DM} \qquad \to \qquad m_{\gamma_D} < 40\;\mathrm{eV} \qquad (\mathrm{dark\; photon \;over\text{-}abundance})
\end{equation}
where we used $g_{\ast s, \rm vis} (T_0) = 3.91$.

\begin{figure}[t]
\centering
\includegraphics[width=0.5\textwidth]{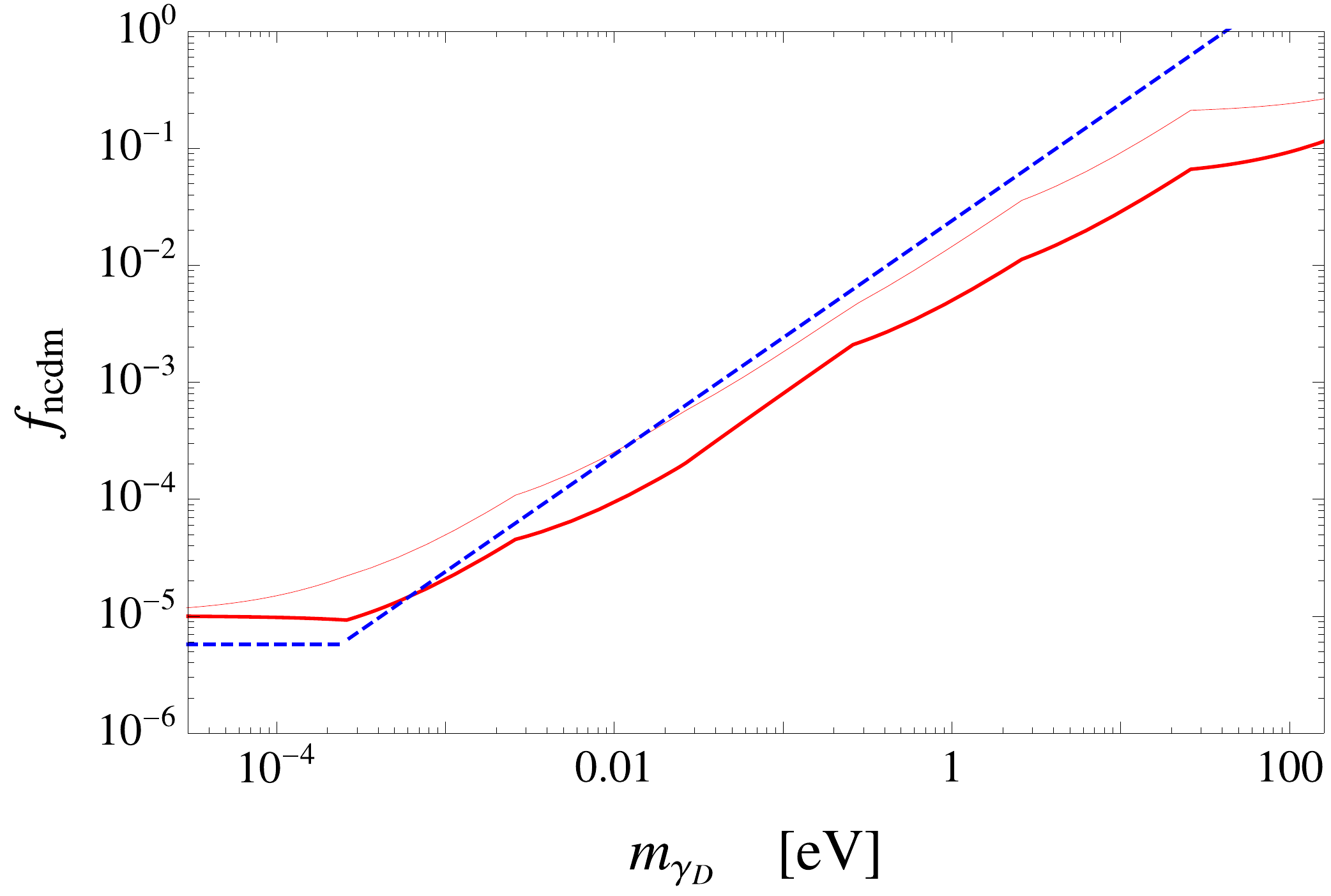}
\caption{The fraction of non-cold DM embodied by the dark photon as predicted by our model (dashed blue), compared to the $2\sigma$ (thick red) and $3\sigma$ (thin red) upper bounds from Ref.~\cite{Diamanti:2017xfo}.}
\label{fig:fncdm}
\end{figure}

Stronger constraints are derived from studies of ``mixed DM'' models, where the DM consists of an admixture of cold and non-cold particles. Recently, Ref.~\cite{Diamanti:2017xfo} obtained bounds on the fraction $f_{\rm ncdm}$ of the non-cold DM component, assumed to be a thermal relic, for a wide range of masses, by combining observations of the CMB, baryon acoustic oscillations (BAO) and the number of dwarf satellite galaxies of the MW. In our model, if the dark photon freezes out when relativistic it constitutes a hot DM component. Its temperature at late times is obtained from entropy conservation, $T_{\gamma_D} / T  = [g_{\ast s, \rm vis} (T) / g_{\ast s, \rm vis} (T_{\rm dec})]^{1/3} \approx 0.37$, where $T$ is the SM photon temperature and we took $ g_{\ast s, \rm vis} (T_{\rm dec}) = 75.75$. The fraction of non-cold DM is
\begin{equation} \label{eq:fncdm_theory}
f_{\rm ncdm} \simeq \frac{\Omega_{\rm ncdm}}{ \Omega_{\rm DM}} = \frac{\rho_{\gamma_D, 0}}{\rho^c_0 \Omega_{\rm DM}} =  \frac{r_{\gamma_D} s_{\rm SM, 0}}{\rho^c_0 \Omega_{\rm DM}}  \begin{cases}
\frac{\pi^4 T_{\gamma_D, 0}}{30\, \zeta (3)}  \\
 m_{\gamma_D}  
\end{cases} \approx
 \begin{cases}
  5.8 \times 10^{-6}  & m_{\gamma_D} \lesssim 3 \,T_{\gamma_D,0} \\
0.024\left( \frac{m_{\gamma_D}}{1\;\mathrm{eV}} \right)   & m_{\gamma_D} \gtrsim 3 \,T_{\gamma_D,0} 
\end{cases}
\end{equation}
where the first (second) expression applies to the case where the dark photon is still relativistic (non-relativistic) today, with $3 \,T_{\gamma_D,0} \approx 2.6 \times 10^{-4}\;\mathrm{eV}$. In the first equality we assumed $\Omega_{\rm ncdm} \ll \Omega_{\rm DM}$ since the non-cold component is in practice constrained to be small, while $ \rho^c = 3 H^2 M_{\rm Pl}^2$ is the critical density. The prediction in Eq.~\eqref{eq:fncdm_theory} can be compared with the bounds given in Ref.~\cite{Diamanti:2017xfo}, after correcting for the fact that there the non-cold relic was assumed to have temperature equal to that of the SM neutrinos, hence the mass needs to be rescaled by a factor $T_{\gamma_D}/T_\nu \approx 0.52$. The result is shown in Fig.~\ref{fig:fncdm}, from which we read a $95\%$ CL bound 
\begin{equation} \label{eq:ncdm_Bound}
m_{\gamma_D} < 6 \times 10^{-4}\;\mathrm{eV}, \qquad (\mathrm{CMB} + \mathrm{BAO} + \mathrm{MW}\;\mathrm{satellites})
\end{equation}
roughly equivalent to the requirement that $\gamma_D$ be still relativistic today. For dark photon masses that satisfy the overclosure bound of Eq.~\eqref{eq:overclosureBound} the relevant observables are CMB and BAO measurements, while the MW satellite count becomes important at higher masses, of order $\mathrm{keV}$ \cite{Diamanti:2017xfo}. In the region $m_{\gamma_D} \lesssim 1\;\mathrm{eV}$, where the dark photon behaved as radiation at photon decoupling, the constraints shown in Fig.~\ref{fig:fncdm} are stronger than those derived purely from $\Delta N_{\rm eff}$. This is due to the inclusion of BAO, which are sensitive to the suppression of the matter power spectrum on small scales caused by the free-streaming of the hot DM component. 

For dark photon masses satisfying Eq.~\eqref{eq:ncdm_Bound}, the phenomenology for $m_{\gamma_D} = 0$ discussed in Sec.~\ref{subsec:masslessDP} still applies. The $\chi$ annihilation is unaffected, including the SE, as the dark photon mediates an effectively long-range force: its wavelength is much larger than the Bohr radius of the $(\chi^\ast \chi)$ bound state, $m_{\gamma_D} \ll \alpha_D m_\chi / 2$. In addition, the Coulomb limit of Eq.~\eqref{eq:SECoulomb} is still appropriate, since the average momentum transfer is much larger than the mediator mass, $m_{\gamma_D} \ll m_{\chi} v_{\rm rel} / 2$ \cite{Petraki:2016cnz}. In the calculation of the ellipticity bound for massless dark photon \cite{Agrawal:2016quu} the infrared divergence that arises from integrating Eq.~\eqref{eq:SIDMxsec} over angles was cut off at the inter-particle distance, $ \lambda_P = ( m_\chi / \rho_\chi )^{1/3} \sim 5\;\mathrm{cm} $, where the numerical value was estimated for a representative DM mass $m_\chi = 100\;\mathrm{GeV}$ and density $\rho_\chi \sim 1 \;\mathrm{GeV}/\mathrm{cm}^3$ in the DM-dominated outer region ($\mathrm{r}\geq 6\;\mathrm{kpc}$) of NGC720 \cite{Humphrey:2010hd}. When $m_{\gamma_D} > 1/\lambda_P \sim 4 \times 10^{-6}\, \mathrm{eV}$, it is $1/m_{\gamma_D}$ that must be taken as IR cutoff. However, since the cutoff only enters logarithmically in the expression of the timescale for velocity isotropization, the ellipticity bound discussed for $m_{\gamma_D} = 0$ applies essentially unchanged to the whole region defined by Eq.~\eqref{eq:ncdm_Bound}. The same holds for the bound from dwarf galaxy survival.\footnote{The small dark photon masses in Eq.~\eqref{eq:ncdm_Bound} are legitimate from an EFT standpoint. Still, it has recently been conjectured~\cite{Reece:2018zvv} that quantum gravity forbids arbitrarily small St\"uckelberg masses: local quantum field theory would break down at $\Lambda_{\rm UV} \sim (m_{\gamma_D} M_{\rm Pl}/g_D)^{1/2}$. Taking $g_D \sim 0.1$ as needed to obtain the observed relic density for $\chi$, Eq.~\eqref{eq:ncdm_Bound} corresponds then to a troublesome $\Lambda_{\rm UV} \lesssim 4\;\mathrm{TeV}$. The conjecture does not apply, however, if $m_{\gamma_D}$ arises from a dynamical symmetry breaking~\cite{Reece:2018zvv}. This topic is currently under debate~\cite{Craig:2018yld}.}

For $ 3 T_{\rm fo}^\chi \approx 3 m_\chi / 25 \lesssim m_{\gamma_D} < m_\chi $ the dark photon freezes out non-relativistically, but is nevertheless over-abundant.

\subsubsection{Heavy dark photon: $m_{\chi} < m_{\gamma_D}$}
\begin{figure}[t]
\centering
\includegraphics[width=0.47\textwidth]{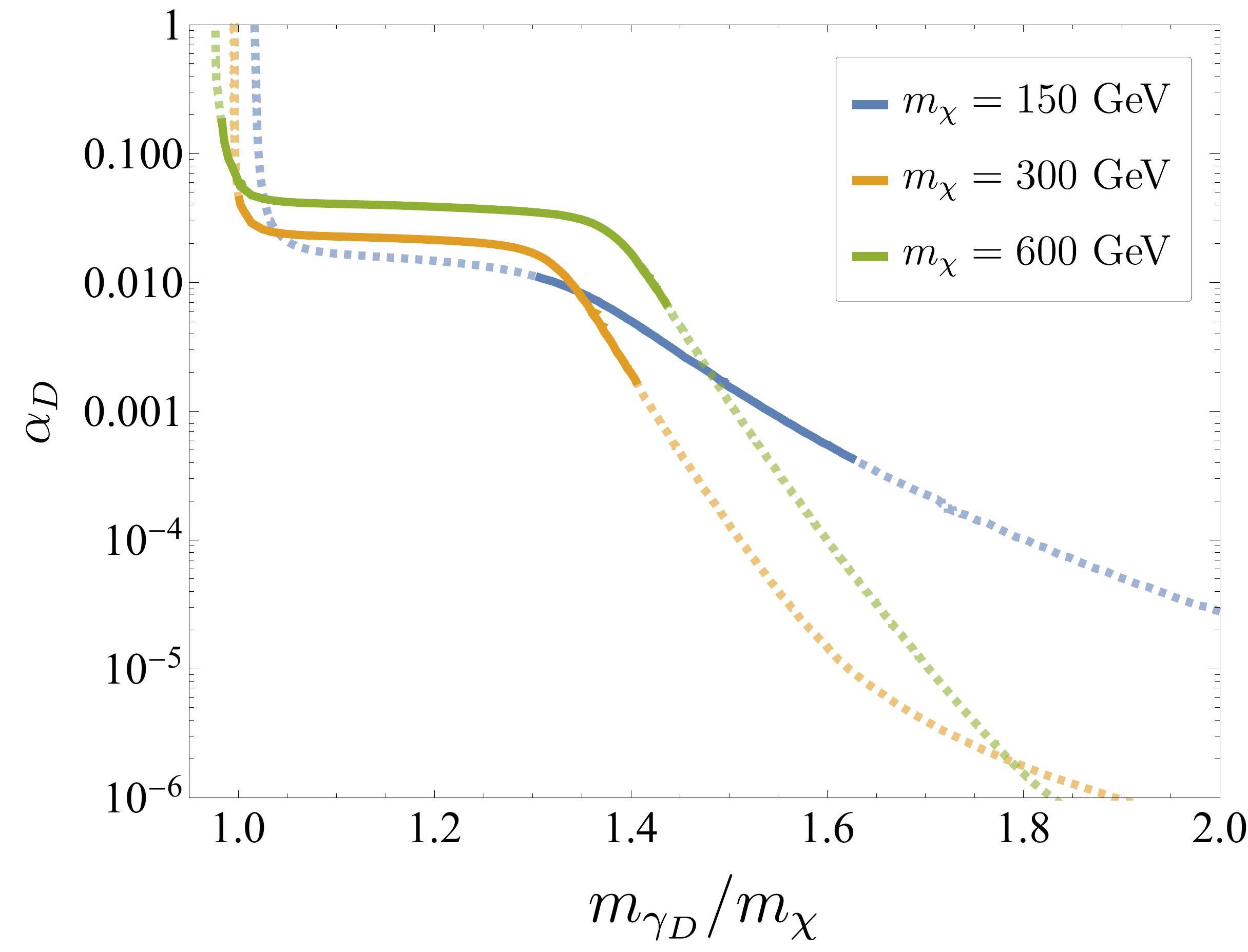} \hspace{1mm}
\includegraphics[width=0.4925\textwidth]{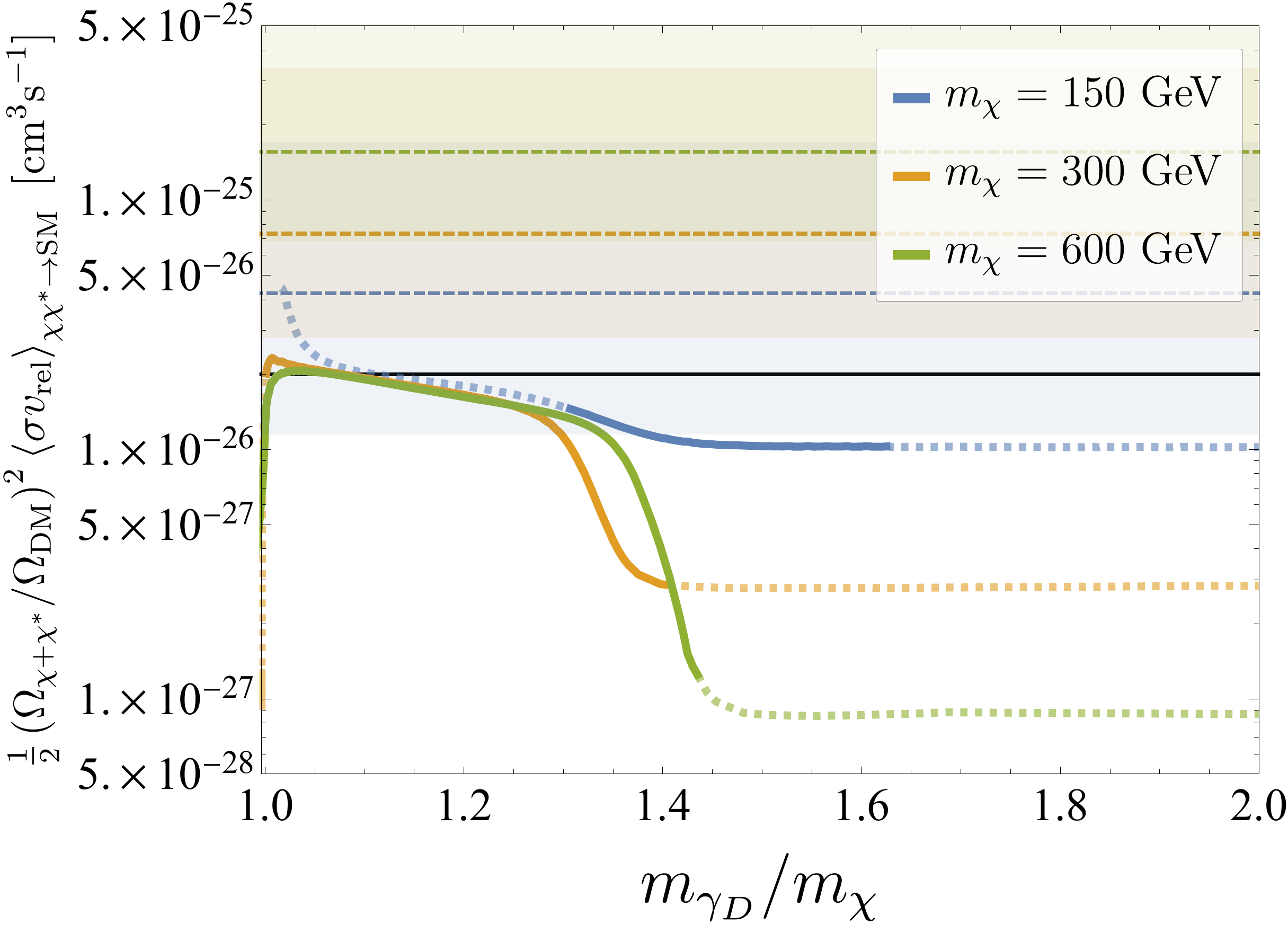}
\caption{{\it Left panel:} contours in the $(m_{\gamma_D}/m_\chi, \alpha_D)$ plane where the sum of the $\chi$ and $\gamma_D$ densities matches the observed total DM density, $\Omega_{\chi + \chi^\ast} + \Omega_{\gamma_D} = \Omega_{\rm DM}$, assuming $f = 1\;\mathrm{TeV}$ and for representative values of $m_\chi > m_\chi^{(f)} \approx 122\;\mathrm{GeV}$. The solid portions highlight the range of $\alpha_D$ where $m_\chi$ can be obtained from dark photon loops cut off at $2.5\;\mathrm{TeV} < m_\rho < 4\pi f$ (see Eq.~\eqref{eq:gaugeDMmass}), where the lower bound comes from the $S$ parameter, $\widehat{S} \sim m_W^2/ m_\rho^2 \lesssim 10^{-3}$ (see e.g. Ref.~\cite{Giudice:2007fh}). {\it Right panel:} effective cross section for present-day DM annihilation to SM particles, calculated along the relic density contours in the left panel. Also shown are the observed $95\%$ CL limits from dSphs in the $WW$ channel \cite{Ackermann:2015zua} (dashed lines), together with the $95\%$ CL uncertainties on the expected limits (colored regions). For reference, the black solid line shows $\langle \sigma v_{\rm rel} \rangle_{\rm can}$, the cross section expected for a single thermal relic that annihilates entirely to SM particles.}
\label{fig:2compMain}
\end{figure}

In the region $m_\chi \lesssim m_{\gamma_D} < 2m_\chi$ both $\gamma_D$ and $\chi$ are stable and freeze out when non-relativistic, naturally giving rise to a two-component cold DM model.
The features of this region are best explained by fixing $f$ and $m_\chi > m_\chi^{(f)}$, so that $\chi$ would be under-abundant in isolation, owing to its too strong annihilation to SM particles via the derivative Higgs portal. Requiring that the heavier dark photon provides the remaining DM fraction then gives a contour in the $(m_{\gamma_D}/m_\chi, \alpha_D)$ plane, shown in the left panel of Fig.~\ref{fig:2compMain} for $f = 1\;\mathrm{TeV}$ and some representative choices of $m_\chi$. The relic densities of $\chi$ and $\gamma_D$ were computed solving the coupled Boltzmann equations with micrOMEGAs \cite{Belanger:2018mqt}. To understand the basic features of Fig.~\ref{fig:2compMain}$\,$-left, a useful first approximation is to treat the freeze-outs of $\chi$ and $\gamma_D$ as decoupled processes, since in this limit the relic density of $\chi$ is simply fixed by the freeze-out of $\chi \chi^\ast \to \mathrm{SM}$ and therefore completely determined by $f$ and $m_{\chi}$. This simplified picture does receive important corrections in some regions of parameter space, as we discuss below. Focusing first on the $m_\chi = 300\;\mathrm{GeV}$ case, four qualitatively different regions arise in our analysis:\enlargethispage{-12mm} 
\begin{enumerate}[label=(\alph*)]
\item The non-degenerate region, $ 2 m_\chi - m_h \approx 1.6\, m_\chi < m_{\gamma_D} < 2 m_\chi $. The dark photon freeze-out is determined by the semi-annihilation process $\gamma_D h \to \chi \chi^\ast$, which is kinematically allowed at zero temperature. Hence the relic density contour is approximately given by $n_h^{\rm eq} \langle \sigma_{\gamma_D h \to \chi \chi^\ast} v_{\rm rel} \rangle = \mathrm{constant} $, where the LHS is evaluated at the $\gamma_D$ freeze-out temperature, $T_{\gamma_D}^{\rm fo} \approx m_{\gamma_D} / 25$, and the thermally averaged cross section is given in Eq.~\eqref{eq:semiannihilationGammah}. As $m_{\gamma_D}/m_\chi$ decreases, the dark fine structure constant increases exponentially to compensate for the suppression of the Higgs number density, $\alpha_D \propto \exp \big(\frac{m_h}{m_\chi} \frac{25}{m_{\gamma_{D}}/m_\chi} \big)$, where we dropped subleading power corrections. The importance of semi-annihilation processes, which change the total DM number by one unit (rather than two units as for ordinary annihilation), was discussed for the first time in Ref.~\cite{DEramo:2010keq}.
\item The intermediate region, $ 1.3\,m_\chi \lesssim m_{\gamma_D} \lesssim  1.6\,m_\chi \approx  2 m_\chi - m_h$. The $\gamma_D$ freeze-out is still determined by $\gamma_D h \to \chi \chi^\ast$, which however is now forbidden at zero temperature. Using detailed balance, the relic density contour is given by $n_h^{\rm eq} \langle \sigma_{\gamma_D h \to \chi \chi^\ast} v_{\rm rel} \rangle = (n_{\chi}^{\mathrm{eq}\,2} / n_{\gamma_D}^{\rm eq})  \langle \sigma_{\chi \chi^\ast \to \gamma_D h} v_{\rm rel} \rangle = \mathrm{constant} $, where the LHS is evaluated at $T_{\gamma_D}^{\rm fo} \approx m_{\gamma_D} / 25$ and the cross section can be found in Eq.~\eqref{eq:semiannihilationChiChi}. The dependence of $\alpha_D$ on $m_{\gamma_D}/m_\chi$ is exponential and faster than in the non-degenerate region, $\alpha_D \propto \exp \big[\big(2 -\tfrac{m_{\gamma_D}}{m_\chi}\big) \frac{25}{m_{\gamma_D}/m_\chi}\big]$, where power corrections were neglected. 
\item The degenerate region, $ m_\chi \lesssim m_{\gamma_D} \lesssim 1.3\,m_\chi $. As $m_{\gamma_D}/m_\chi$ decreases the semi-annihilation is increasingly Boltzmann suppressed, while the rate of the annihilation $\gamma_D \gamma_D \to \chi \chi^\ast$ increases as $\alpha_D^2\,$. Therefore the dark photon freezes out when its annihilation to $\chi \chi^\ast$ goes out of equilibrium. The relic density contour is approximately described by $\langle \sigma_{\gamma_D \gamma_D \to \chi \chi^\ast} v_{\rm rel} \rangle = \mathrm{constant}$, where the cross section is given in Eq.~\eqref{eq:annihilationChiChi}. The resulting variation of $\alpha_D$ is slow in comparison to the regions dominated by semi-annihilation, thus explaining the nearly flat behavior of the contours. Importantly, in this region the evolutions of the $\chi$ and $\gamma_D$ densities are tightly coupled, and the injection of $\chi$ particles due to the $\gamma_D \gamma_D \to \chi \chi^\ast$ process gives a larger $\chi$ abundance than the one expected based on the simplified decoupled picture. This interesting type of system was first studied numerically in Ref.~\cite{Belanger:2011ww}, and we provide here analytical insight into its dynamics. After the yields $Y_{\chi, \gamma_D}$ become much larger than their equilibrium values, they obey the simplified Boltzmann equations ($x \equiv m_\chi / T$)
\begin{subequations}
\begin{alignat}{2}
\widehat{\lambda}^{-1} x^2\, \frac{dY_\chi}{dx} &\,=\, - \langle \sigma v_{\rm rel} \rangle_{\mathrm{SM}} Y_\chi^2 + \tfrac{1}{2} \langle \sigma v_{\rm rel} \rangle_{\gamma_D \gamma_D} Y_{\gamma_D}^2 \\
\widehat{\lambda}^{-1} x^2\, \frac{dY_{\gamma_D}}{dx} &\,=\, - \langle \sigma v_{\rm rel} \rangle_{\gamma_D \gamma_D} Y_{\gamma_D}^2
\end{alignat}
\end{subequations}
where $\widehat{\lambda} \equiv (2\sqrt{10}\, \pi / 15) (g_{\ast s} m_\chi M_{\rm Pl} / \sqrt{g_\ast})$, while $\langle \sigma v_{\rm rel} \rangle_{\mathrm{SM}}$ refers to $\chi \chi^\ast \to \mathrm{SM}$ and $\langle \sigma v_{\rm rel} \rangle_{\gamma_D \gamma_D}$ to $\gamma_D \gamma_D \to \chi \chi^\ast$. The analytical solution of this system gives at $x \gg 1$
\begin{equation} \label{eq:2CDMsolution}
\frac{1}{a_\sigma}\left(\frac{2Y_{\chi}}{Y_{\gamma_D}}\right)^2 \simeq  1 + \tfrac{1}{2}\left( a_\sigma + \sqrt{a_\sigma (a_\sigma + 4)}\, \right), \qquad a_\sigma \equiv \frac{\langle \sigma v_{\rm rel} \rangle_{\gamma_D \gamma_D}}{\langle \sigma v_{\rm rel} \rangle_{\mathrm{SM}}/2}\;,
\end{equation}
where $a_\sigma$ goes to a constant since both processes are $s$-wave. This result is obtained by solving a quadratic equation, whose other root yields $dY_\chi / dx > 0$ and is therefore unphysical. For $a_\sigma \ll 1$, as verified in the $m_\chi = 300, 600$ GeV examples, the RHS of Eq.~\eqref{eq:2CDMsolution} goes to $1$ and the formula expresses the equality of the fluxes that enter and leave the $\chi$ population, $Y_{\gamma_D}^2 \langle \sigma v_{\rm rel} \rangle_{\gamma_D \gamma_D} = (2Y_\chi)^2 \langle \sigma v_{\rm rel} \rangle_{\mathrm{SM}}/2\,$. Correspondingly, the relative $\chi$ density is suppressed (albeit still larger than in the simplified decoupled picture), $2n_\chi / n_{\gamma_D} \simeq a_\sigma^{1/2}$. In the $m_\chi = 150\;\mathrm{GeV}$ example we have $a_\sigma = O(1)$ instead: in this regime the annihilation to the SM is not as efficient, leading to an accumulation of the $\chi$ particles injected by $\gamma_D\gamma_D$ annihilation and therefore to a large relative $\chi$ abundance, $2n_\chi / n_{\gamma_D} \simeq \mathrm{few}$.

\item The very degenerate and forbidden~\cite{Griest:1990kh} region, $m_{\gamma_D} \lesssim m_\chi $. The dark photon freeze-out is determined by $\gamma_D \gamma_D \to \chi \chi^\ast$, but $\alpha_D$ increases very rapidly as $m_{\gamma_D}/m_\chi$ is decreased toward and eventually slightly below $1$, in order to compensate for the kinematic suppression.
\end{enumerate}  
The previous discussion focused on the $m_{\chi} = 300\;\mathrm{GeV}$ benchmark. The features of the relic density contour for $m_{\chi} = 600\;\mathrm{GeV}$ are very similar. On the contrary, in the case $m_\chi = 150\;\mathrm{GeV}$ we have $2m_\chi - m_{h} \approx 1.2\, m_\chi$ and as a consequence we observe a direct transition from the non-degenerate to the degenerate region, while the intermediate region is absent.

The right panel of Fig.~\ref{fig:2compMain} shows the effective cross section for DM annihilation to SM particles today, computed along the relic density contours. All processes that yield SM particles were included in the numerical evaluation, but we have checked that $\chi \chi^\ast \to \mathrm{SM}$ is always dominant and the subleading channels (such as $\gamma_D \chi \to h \chi$ and $\gamma_D \gamma_D \to \mathrm{SM}$, the latter of which proceeds at one loop) contribute at the sub-percent level.\footnote{Note that due to the large mass of the dark photon, in this case the Sommerfeld enhancement of the $\chi \chi^\ast \to \mathrm{SM}$ annihilation is negligible.} Two different regimes can be observed. In the non-degenerate region the freeze-outs of $\chi$ and $\gamma_D$ can be treated as independent to a good approximation, hence from Eq.~\eqref{eq:RAapproxAnalytical} the effective cross section is reduced compared to the standard thermal value $ \langle \sigma v_{\rm rel} \rangle_{\rm can} \approx 2 \times 10^{-26} \,\mathrm{cm}^3\, \mathrm{s}^{-1}$ by a factor $\langle \sigma v_{\rm rel} \rangle_{\rm can} \,/ (\tfrac{1}{2}\langle \sigma v_{\rm rel} \rangle_{\chi \chi^\ast \to\, \mathrm{SM}} ) < 1$. For $m_\chi = 600 \;\mathrm{GeV}$ the suppression amounts to more than one order of magnitude. Conversely, in the degenerate region the already discussed injection of $\chi$ particles from $\gamma_D \gamma_D$ annihilations compensates the increased $\langle \sigma v_{\rm rel} \rangle_{\chi \chi^\ast \to\, \mathrm{SM}}\,$, resulting in effective cross sections that are numerically close to $\langle \sigma v_{\rm rel} \rangle_{\rm can}$.

Finally, if $ 2 m_\chi < m_{\gamma_D}$ the dark photon is unstable, with decay width $\Gamma(\gamma_D \to \chi^\ast \chi) = (\alpha_D m_{\gamma_D} / 12) (1 - 4m_\chi^2/m_{\gamma_D}^2)^{3/2}$. In the early Universe, the inverse decay process keeps the dark sector in chemical equilibrium until $H \sim \langle \Gamma \rangle \, n_{\gamma_D} / n_\chi$, when the ratio of the number densities is
\begin{equation}
\frac{n_{\gamma_D}}{n_{\chi}} \sim \frac{H}{\Gamma} \sim \frac{10\, T^2}{M_{\rm Pl} \alpha_D m_{\gamma_D}} < 10^{-12}\, \left( \frac{m_{\gamma_D}}{100\;\mathrm{GeV}} \right) \left( \frac{10^{-3}}{\alpha_D} \right),
\end{equation}
where we assumed that $T < m_{\gamma_D}$ at this point, and neglected $O(1)$ factors. Thus, the subsequent decay of the remaining dark photons has negligible impact on the $\chi$ relic density, which can effectively be computed considering only the freeze-out of $\chi\chi^\ast$ annihilations to SM particles, with the results summarized in Fig.~\ref{fig:fvsmchi}. In the region $2 m_\chi < m_{\gamma_D}$ the only phenomenologically relevant imprint of the dark photon is the one-loop mass for $\chi$, estimated in Eq.~\eqref{eq:potentialGauging}.

\vspace{0.5cm}
\section{Closing remarks} \label{sec:outlook}
\noindent We have considered models where the Higgs doublet $H$ and the DM $\chi$ have common origin as pNGBs of a spontaneously broken global symmetry. We have shown that the shift symmetry of $\chi$ can be broken in such ways that a mass of $O(100)\;\mathrm{GeV}$ is generated at one loop, whereas the non-derivative couplings between $\chi$ and the SM are small, naturally leading to suppressed direct detection. In a first realization the DM, taken to be either a real or complex scalar, acquires mass from bottom quark loops. Correspondingly the operator $y_b \bar{q}_L H b_R\, |\chi|^2/f^2$ is generated with $O(1)$ coefficient, leading to very suppressed cross sections for DM-nucleus scattering that will be probed by LZ. In a second realization, which constitutes the central subject of this work, the DM is a complex scalar whose mass arises from the gauging of the $U(1)_{\rm DM}$ stabilizing symmetry. The direct detection signal is out of reach even at future experiments, but the dark sector -- now including $\chi$ and the dark photon $\gamma_D$ as light fields -- can be tested both at colliders and in cosmology and astroparticle experiments. As concerns the latter, especially important observables are $\Delta N_{\rm eff}$ and the effects of long-range DM self-interactions if $m_{\gamma_D} = 0$, and indirect DM detection if $m_{\gamma_D} > m_\chi\,$.

We wish to remark that promoting $U(1)_{\rm DM}$ to a local symmetry may in fact be preferred, based on both model-specific and more general theoretical considerations. Specifically, gauging $U(1)_{\rm DM}$ ensures that any subleading couplings of the SM fermions to the strong sector, which were neglected in our discussion, automatically preserve the DM stability.\footnote{We thank K.~Agashe and M.~Frigerio for enlightening comments on this point.} More generally, several arguments exist that suggest quantum gravity does not conserve continuous global symmetries (see Refs.~\cite{Kallosh:1995hi,ArkaniHamed:2006dz,Banks:2006mm} and further references therein). If that is the case, then Planck-scale suppressed operators can destabilize the DM, potentially leading to conflict with observations~\cite{Mambrini:2015sia}, although this strongly depends on the assumptions made about the coefficients of the higher-dimensional operators. In any case, the issue is absent if $U(1)_{\rm DM}$ is gauged.

We conclude with some further comments about the collider phenomenology, focusing on signals that involve the pNGB DM (overviews of the ``standard'' signatures of composite Higgs models can be found in Refs.~\cite{Panico:2015jxa,Bellazzini:2014yua}). As already discussed, for $m_\chi < m_h/2$ the searches for invisible Higgs decays provide a powerful probe of the derivative Higgs portal operator. In contrast, the experimental prospects are less favorable for $m_\chi > m_h/2$, when the intermediate Higgs is off shell. For the marginal Higgs portal the reach was studied in Ref.~\cite{Craig:2014lda} for the LHC and future hadron colliders, including all relevant production channels (monojet, $t\bar{t}h$ and vector boson fusion), and in Ref.~\cite{Chacko:2013lna} for future lepton colliders. For real DM with $m_\chi = 100\;\mathrm{GeV}$ the sensitivity was found to extend up to $\lambda \sim 0.7$ at the High-Luminosity LHC and $\lambda \sim 0.3$ at a $100\;\mathrm{TeV}$ $pp$ collider with $30$ ab$^{-1}$ (neglecting systematic uncertainties~\cite{Craig:2014lda}), and $\lambda \sim 0.6$ at the $1\;\mathrm{TeV}$ ILC \cite{Chacko:2013lna}. For the derivative Higgs portal relevant to pNGB DM, we need to replace $\lambda \to c_d M^2_{\chi \chi^\ast}/(2f^2)$ where $M^2_{\chi \chi^\ast} \geq 4m_\chi^2$ is the squared invariant mass of the DM pair. Although the momentum-dependent coupling gives harder kinematic distributions and therefore sensitivity to smaller cross sections compared to the momentum-independent case, the LHC reach in the monojet channel is negligible~\cite{Barducci:2016fue}. A detailed assessment of the reach of future colliders on the derivative portal is, to our knowledge, not yet available. Another class of signals arises from composite resonances that are charged under the DM-stabilizing symmetry. For fermionic top partners, the reach at future hadron colliders was shown to exceed that on resonances with only SM quantum numbers~\cite{Chala:2018qdf}. For vector resonances, the pair production (via Drell-Yan and vector boson fusion) and the production in association with DM yield final states with $W$ and/or $Z$ bosons, missing transverse energy, and possibly Higgs bosons. The latter signatures, although difficult to discover due to the suppressed cross sections, constitute a robust feature of models where the Higgs and the DM arise as pNGB, regardless of the details of the specific construction.

\vspace{1cm}
\noindent {\bf Acknowledgments} 

\noindent We have benefited from conversations with K.~Agashe, M.~Frigerio, M.~Garny, R.~Harnik, A.~Katz, J.~Serra, Y.~Shadmi, and Y.~Tsai. We also thank P.~Agrawal for clarifications about Ref.~\cite{Agrawal:2016quu}. This work has been partially supported by the DFG Cluster of Excellence 153 ``Origin and Structure of the Universe,'' by the Collaborative Research Center SFB1258, and the COST Action CA15108. RB is supported by the Minerva foundation. MR is supported by the Studienstiftung des deutschen Volkes. The work of ES was initiated at the Aspen Center for Physics, which is supported by NSF grant PHY-1607611. ES thanks the organizers of the CERN-Korea TH Institute ``Physics at the LHC and beyond'' for a stimulating environment and partial support in the final stages of this work. ES is also grateful to the GGI for hospitality and to the INFN for partial support as this paper was being completed.
\appendix
\section{$SO(7)/SO(6)$ model: fermion sector} \label{sec:appA}
The Callan-Coleman-Wess-Zumino (CCWZ) construction~\cite{Coleman:1969sm,Callan:1969sn} for the $SO(7)/SO(6)$ coset was given in Appendix~A of Ref.~\cite{Balkin:2017aep}, of which we adopt the notation and conventions. The choice of DM shift symmetry-preserving top quark embeddings that we have made in this paper is
\begin{equation} \label{eq:TopEmbeddings}
\mathbf{7}_{2/3} \sim \xi_L^{(t)} = \frac{1}{\sqrt{2}} \begin{pmatrix} i b_L, & b_L , & i t_L, & - t_L, & \mathbf{0}_3^{\,T} \end{pmatrix}^T, \qquad \mathbf{21}_{2/3} \sim \xi_R^{(t)} = \frac{i\, t_R}{2} \begin{pmatrix} \begin{matrix} 0 & -1 \\  1 & 0 \end{matrix} & & \\  & \begin{matrix} 0 & 1 \\  -1 & 0 \end{matrix} &  \\  & &   \mathbf{0}_{3\times 3} \end{pmatrix} ,
\end{equation}
where empty entries in the expression of $\xi_R^{(t)}$ are zeros. Since $\mathbf{7} = \mathbf{6} \,\oplus\, \mathbf{1}$ and $\mathbf{21}= \mathbf{15} \,\oplus \, \mathbf{6}$ under $SO(6)$, in the top sector we expect fermionic resonance multiplets $G \sim\mathbf{15}_{2/3}, \,Q \sim \mathbf{6}_{2/3}$ and $S \sim \mathbf{1}_{2/3}$ under $SO(6)\times U(1)_X$. The decomposition and component expression of $Q$ was given in Sec.~II of Ref.~\cite{Balkin:2017aep}, whereas $G$ decomposes as $\mathbf{15} = [(\mathbf{3},\mathbf{1}) + (\mathbf{1},\mathbf{3})]_{0} \oplus (\mathbf{1}, \mathbf{1})_0 \oplus (\mathbf{2}, \mathbf{2})_{\pm 1}$ under $SU(2)_L \times SU(2)_R \times U(1)_{\rm DM}$, where the $X = 2/3$ charge is understood. In components,
\begin{equation}
G = \frac{1}{2} \begin{pmatrix} 0 & -i T^{12}_+ & B^{12}_{+} - X^{12}_{+} & -i ( B^{12}_- + X^{12}_-) & - \mathcal{B}_{-} - \mathcal{X}_{5/3\,-} & - i ( \mathcal{B}_{+} + \mathcal{X}_{5/3\,+} )  \\
& 0 & -i (B^{12}_+ + X^{12}_+) & - B^{12}_- + X^{12}_- & i ( \mathcal{B}_- - \mathcal{X}_{5/3\,-}) & -  \mathcal{B}_{+} + \mathcal{X}_{5/3\,+}   \\
& & 0 & -i T^{12}_- & - \mathcal{T}_-  - \mathcal{X}_{2/3\, -} &  - i ( \mathcal{T}_+  + \mathcal{X}_{2/3\, +} ) \\
& & & 0 & - i ( \mathcal{T}_-  - \mathcal{X}_{2/3\, -}) &  \mathcal{T}_{+}  - \mathcal{X}_{2/3\, +} \\
& & & & 0 & -i \sqrt{2}\, \widetilde{S} \\
& & & & & 0 \end{pmatrix} , 
\end{equation}
where the lower triangle is determined by antisymmetry. We have made the definitions $T^{12}_\pm \equiv \widetilde{T}_1 \pm \widetilde{T}_2$, $Q^{12}_\pm \equiv (\widetilde{Q}_1 \pm \widetilde{Q}_2)/\sqrt{2}$ ($Q = X, B$) and $\mathcal{Q}_\pm \equiv (\mathcal{Q}^{(+)} \pm \mathcal{Q}^{(-)})/\sqrt{2}$ ($\mathcal{Q} = \mathcal{T}, \mathcal{B}, \mathcal{X}_{2/3}, \mathcal{X}_{5/3}$). Here $(\widetilde{X}_i, \widetilde{T}_i, \widetilde{B}_i )^T$ is a $(\mathbf{3}, \mathbf{1})_0$ for $i =1$ and a $(\mathbf{1}, \mathbf{3})_0$ for $i =2$, $\widetilde{S}\sim (\mathbf{1}, \mathbf{1})_0\,$, and the fields with calligraphic names compose the $(\mathbf{2}, \mathbf{2})_{\pm 1}$.\footnote{Fields with calligraphic names have the same $SO(4)$ quantum numbers as their non-calligraphic versions. For example $\mathcal{X}_{5/3}^{(\pm)}$ transforms as $X_{5/3}$ under $SO(4)$, but has in addition charge $\pm 1$ under $U(1)_{\rm DM}$.} The elementary-composite mixing Lagrangian for the top sector is 
\begin{align}
\mathcal{L}_{\rm mix}^{(t)} =\, \epsilon^i_{qS} \bar{\xi}_L^{(t) A} U_{A7} S_{R, i} + \epsilon^j_{qQ} \bar{\xi}_L^{(t) A} U_{Aa}  Q^{a}_{R, j} + \epsilon^j_{tQ} \bar{\xi}_R^{(t) BA} \,& U_{Aa}   U_{B7} Q^{a}_{L, j}  \nonumber \\  &+ \epsilon^k_{tG} \bar{\xi}_R^{(t) BA} U_{Aa} U_{Bb} G^{ ab}_{L, k} + \mathrm{h.c.}, 
\end{align}
where repeated indices are summed. Here $\{i, j, k\}$ count the multiplicities of resonances and therefore run from $1$ to $\{N_{S}, N_{Q}, N_{G} \}$, respectively, while $A,B$ are $SO(7)$ indices and $a,b$ are $SO(6)$ indices. Calculability of the one-loop scalar potential is obtained via generalized Weinberg sum rules (WSRs), see Ref.~\cite{Balkin:2017aep} for more details. The minimal field content that gives a completely ultraviolet (UV)-finite one-loop Higgs potential is $N_{S} =  N_{Q} = N_{G} = 1$, which we adopt. The embeddings in Eq.~\eqref{eq:TopEmbeddings} yield a Higgs potential with ``double tuning'' structure \cite{Panico:2012uw}, where parametrically $\Delta^{-1} \sim (v^2/f^2) (\epsilon^t)^2$. 

We now describe the embeddings of the bottom quark in the two models discussed in the main text: the one of Sec.~\ref{eq:bRbreaking}, where the $\chi$ shift symmetry is broken by $b_R$, and the one of Sec.~\ref{sec:gaugeBreaking}, where the $\chi$ shift symmetry is preserved by the bottom sector.

{\bf DM shift symmetry broken by $b$ quark} The bottom quark embeddings are
\begin{equation}
\mathbf{7}_{-1/3} \sim \xi_L^{(b)} = \frac{1}{\sqrt{2}} \begin{pmatrix} - i t_L, & t_L , & i b_L, & b_L, & \mathbf{0}_3^{\,T} \end{pmatrix}^T, \qquad \mathbf{7}_{-1/3} \sim \xi_R^{(b)} = b_R \begin{pmatrix} \mathbf{0}_6^{\,T}, & 1 \end{pmatrix} ^T .
\end{equation}
We thus expect resonances $Q^{(b)} \sim \mathbf{6}_{-1/3}$ and $S^{(b)} \sim \mathbf{1}_{-1/3}$ under $SO(6)\times U(1)_X$. The component expression of $Q^{(b)}$ is
\begin{equation}
Q^{(b)} = \frac{1}{\sqrt{2}} \begin{pmatrix} i U_{-4/3} - i \widetilde{T}, U_{-4/3} + \widetilde{T} , i U_{-1/3} + i \widetilde{B} , - U_{-1/3} + \widetilde{B} , - i \mathcal{V} + i \mathcal{W} , \mathcal{V} +  \mathcal{W}  \end{pmatrix}^T ,
\end{equation}
where under $(SU(2)_L)^{\rm DM}_{Y}$ we have $(U_{-1/3}, U_{-4/3})^T \sim \mathbf{2}_{-5/6}^0$, $(\widetilde{T}, \widetilde{B})^T \sim \mathbf{2}_{1/6}^0$ and $\mathcal{V, W} \sim \mathbf{1}_{-1/3}^{\pm 1}$. The elementary-composite mixing Lagrangian for the bottom sector reads
\begin{equation}
\mathcal{L}_{\rm mix}^{(b)} =\, ( \epsilon^m_{qS^{(b)}} \bar{\xi}_L^{(b) A}  S^{(b)}_{R, m} + \epsilon^m_{bS^{(b)}} \bar{\xi}_R^{(b) A}  S^{(b)}_{L, m} ) U_{A7} + ( \epsilon^n_{qQ^{(b)}} \bar{\xi}_L^{(b) A}  Q^{(b) a}_{R, n}  + \epsilon^n_{bQ^{(b)}} \bar{\xi}_R^{(b) A}  Q^{(b) a}_{L, n} ) U_{Aa} + \mathrm{h.c.}, 
\end{equation}
where $\{m, n\}$ run from $1$ to $\{N_{S^{(b)}}, N_{Q^{(b)}} \}$, respectively. The complete fermionic Lagrangian is $\mathcal{L}_f = (\mathrm{kin.\;terms}) + (\mathrm{resonance\;masses}) + \mathcal{L}_{\rm mix}^{(t)} +  \mathcal{L}_{\rm mix}^{(b)} $, where the kinetic terms include both those for the elementary fields and the CCWZ ones for the resonances. Integrating out the resonances we obtain an effective Lagrangian for the top and bottom quarks and the GBs, which we use to calculate the one-loop potential for $\tilde{h}$ and $\chi$. In particular, for the DM mass parameter we find
\begin{equation}
\quad\qquad\mu^2_{\rm DM} = - \frac{N_c}{4\pi^2f^2} \int_0^\infty dp^2 p^2 \frac{\Pi_{R_1}^b}{\Pi_{R_0}^b}\,, \qquad\qquad (b_R \;\mathrm{loops})
\end{equation}
with Euclidean form factors
\begin{equation} \label{eq:bFormFactors}
\Pi_{R_0}^b = 1 + \sum_{m = 1}^{N_{S^{(b)}}} \frac{|\epsilon_{bS^{(b)}}^m |^2}{p^2 + m^2_{S^{(b)}_m}}\,, \qquad \Pi_{R_1}^b = \sum_{n= 1}^{N_{Q^{(b)}}} \frac{|\epsilon_{bQ^{(b)}}^n |^2}{p^2 + m^2_{Q^{(b)}_n}} - \sum_{m = 1}^{N_{S^{(b)}}} \frac{|\epsilon_{bS^{(b)}}^m |^2}{p^2 + m^2_{S^{(b)}_m}} \,.
\end{equation}
We introduce $N_{Q^{(b)}} = N_{S^{(b)}} = 1$ resonances and to obtain partial calculability of the bottom-induced potential we impose one set of WSRs, which make the dimensionless couplings UV-finite and reduce to logarithmic the degree of divergence of the mass parameters. The WSRs correspond to the relations $\epsilon_{qS^{(b)}}^2 = \epsilon_{qQ^{(b)}}^2$ and $\epsilon_{bS^{(b)}}^2 = \epsilon_{bQ^{(b)}}^2\,$, the latter of which implies from Eq.~\eqref{eq:bFormFactors} that $\mu^2_{\rm DM}$ vanishes for $m^2_{Q^{(b)}} = m^2_{S^{(b)}}$. Assuming $m_{Q^{(b)}}, m_{S^{(b)}} > 0$ we take as solutions to the sum rules $\epsilon_{qS^{(b)}} = +\, \epsilon_{qQ^{(b)}}$ and $\epsilon_{bS^{(b)}} = -\, \epsilon_{bQ^{(b)}}\,$, in which case $\lambda$ does not vanish even for $m_{Q^{(b)}} = m_{S^{(b)}}$. We have then the parametric scalings
\begin{equation} \label{eq:VbBreaking}
\mu_{\rm DM}^2  \simeq  a\,\frac{ N_c}{16\pi^2 } \frac{M_{\ast b}^4}{f^2} (\epsilon_R^b)^2 \,C_{QS} \,, \quad C_{QS} \equiv \frac{m_{Q^{(b)}}^2 - m_{S^{(b)}}^2}{M_{\ast b}^2}\,,  \qquad\quad \lambda  \simeq b\, \frac{N_c}{16\pi^2} \frac{M_{\ast b}^2}{f^2} \, y_b^2 ,
\end{equation}
where $a,b > 0$ are $O(1)$ coefficients and $M_{\ast b}$, defined via Eq.~\eqref{eq:yukawa}, is identified with $M_{\ast b}  = m_{Q^{(b)}} + m_{S^{(b)}}$, which implies $|C_{QS}| < 1$. An important constraint on this setup comes from tree-level corrections to the $Z\bar{b}_L b_L$ coupling, since the embedding of $b_L$ in a $ (\mathbf{2}, \mathbf{2})_{-1/3}$ of $SU(2)_L \times SU(2)_R \times U(1)_X$ is not invariant under the $P_{LR}$ custodial symmetry \cite{Agashe:2006at}. The corrections scale as
\begin{equation}
\frac{g}{c_w} Z_\mu \bar{b}_L \gamma^\mu (g_{b_L}^{\rm SM} + \delta g_{b_L} ) b_L\,, \qquad \delta g_{b_L} \simeq +  (\epsilon_L^b)^2\, \frac{v^2}{f^2}
\end{equation}
($g_{b_L}^{\rm SM} = -1/2 + s_w^2/3$), where the sign is fixed to be positive. For comparison, the experimental bound is $-1.7  < 10^3\, \delta g_{b_L} < +1.4$ at $99\%$ CL \cite{Ghosh:2015wiz}.\footnote{In this model $b_R$ is embedded in a $(\mathbf{1}, \mathbf{1})_{-1/3} \subset \mathbf{7}_{-1/3}$, so the $Z\bar{b}_R b_R$ coupling is protected by $P_{LR}$ and very suppressed. Therefore it makes sense to set $\delta g_{b_R} = 0$ in the electroweak fit. Since $\delta g_{b_L}$ is weakly correlated with the remaining precision observables, we can then simply quote its one-parameter bound.\vspace{1.5mm}} A large $b_L$ compositeness, namely $\epsilon_L^b \sim 1$ and $\epsilon_R^b \sim y_b f / M_{\ast b}$, leads to $\mu_{\rm DM}^2 \lesssim \lambda v^2$ and therefore very light DM, $m_\chi \approx 4\;\mathrm{GeV}\, (M_{\ast b}/8\;\mathrm{TeV})(1\;\mathrm{TeV}/f)$,\footnote{We have fixed the numerical value of $y_b$ via $m_b = m_b^{\overline{MS}} (2\;\mathrm{TeV}) \simeq 2.5 \;\mathrm{GeV}$.} but is robustly ruled out by $Z\bar{b}_L b_{L}$ unless $f \gg \mathrm{TeV}$. Conversely, a large $b_R$ compositeness $\epsilon_R^b \sim 1$, $\epsilon_L^b \sim y_b f / M_{\ast b}$ easily satisfies the $Z\bar{b}_L b_L$ constraint. This region, however, yields parametric scalings for $\mu^2_{\rm DM}$ and $\lambda$ that are similar to those already discussed in the case where the DM shift symmetry is broken by $t_R$ couplings. We are thus led to focus on the ``intermediate'' range $\epsilon_L^b \sim \epsilon_R^b \sim \sqrt{y_b f / M_{\ast b}}\,$, where the correction to $Z\bar{b}_L b_L$ is typically moderate, $10^3\,\delta g_{b_L} \sim + \,\mathrm{few} \times 0.1 \left( 8\;\mathrm{TeV} / M_{\ast b} \right) \left( 1\;\mathrm{TeV}/f \right)$, and the DM potential scales as in Eq.~\eqref{eq:parametersbBreaking}, where in the (crude) estimate of the DM mass we have taken a typical $C_{QS} \sim 0.2$ for this parameter region.

\begin{figure}[t]
\centering
\includegraphics[width=.49\textwidth]{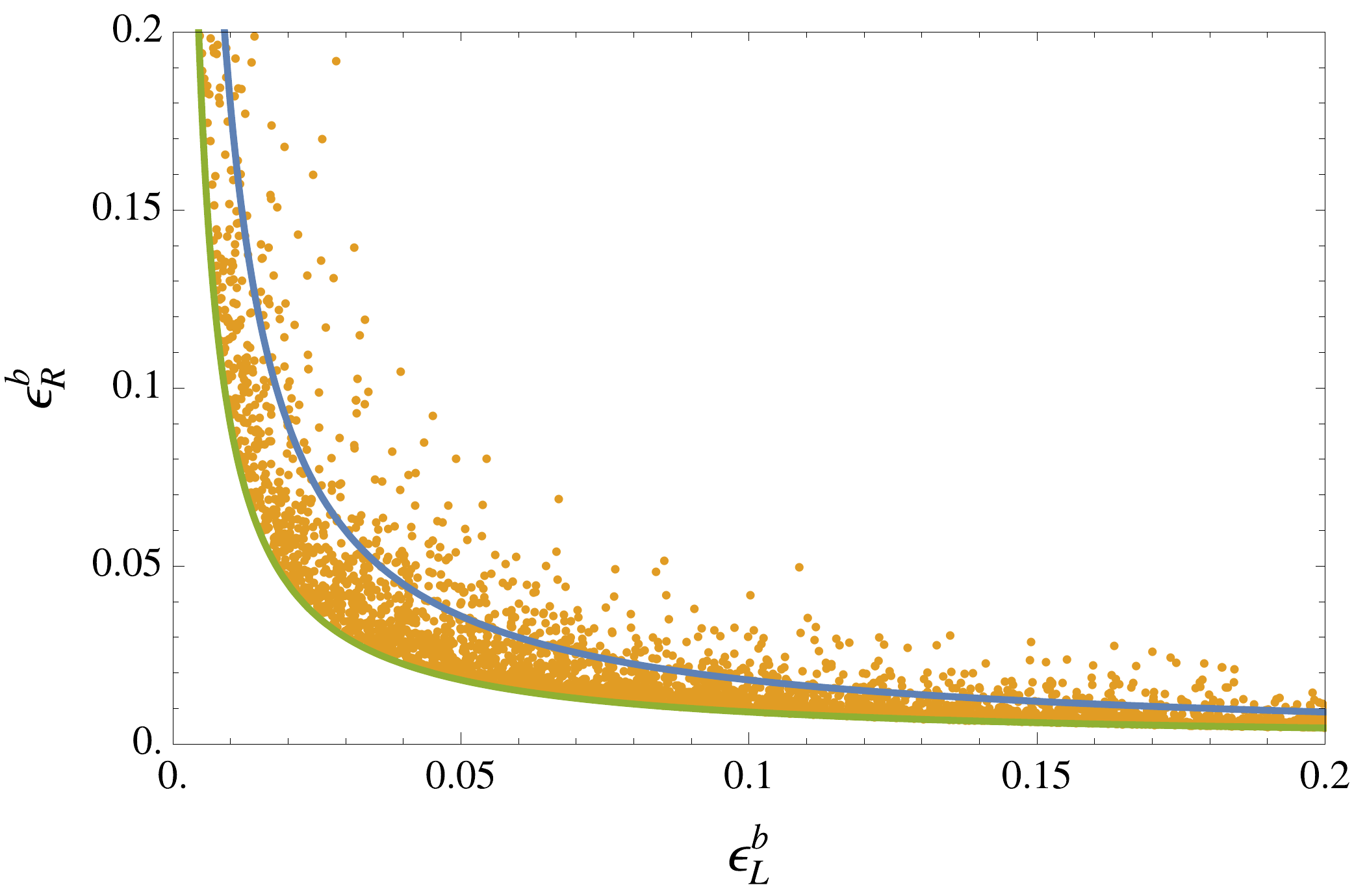} \hspace{1mm}
\includegraphics[width=.49\textwidth]{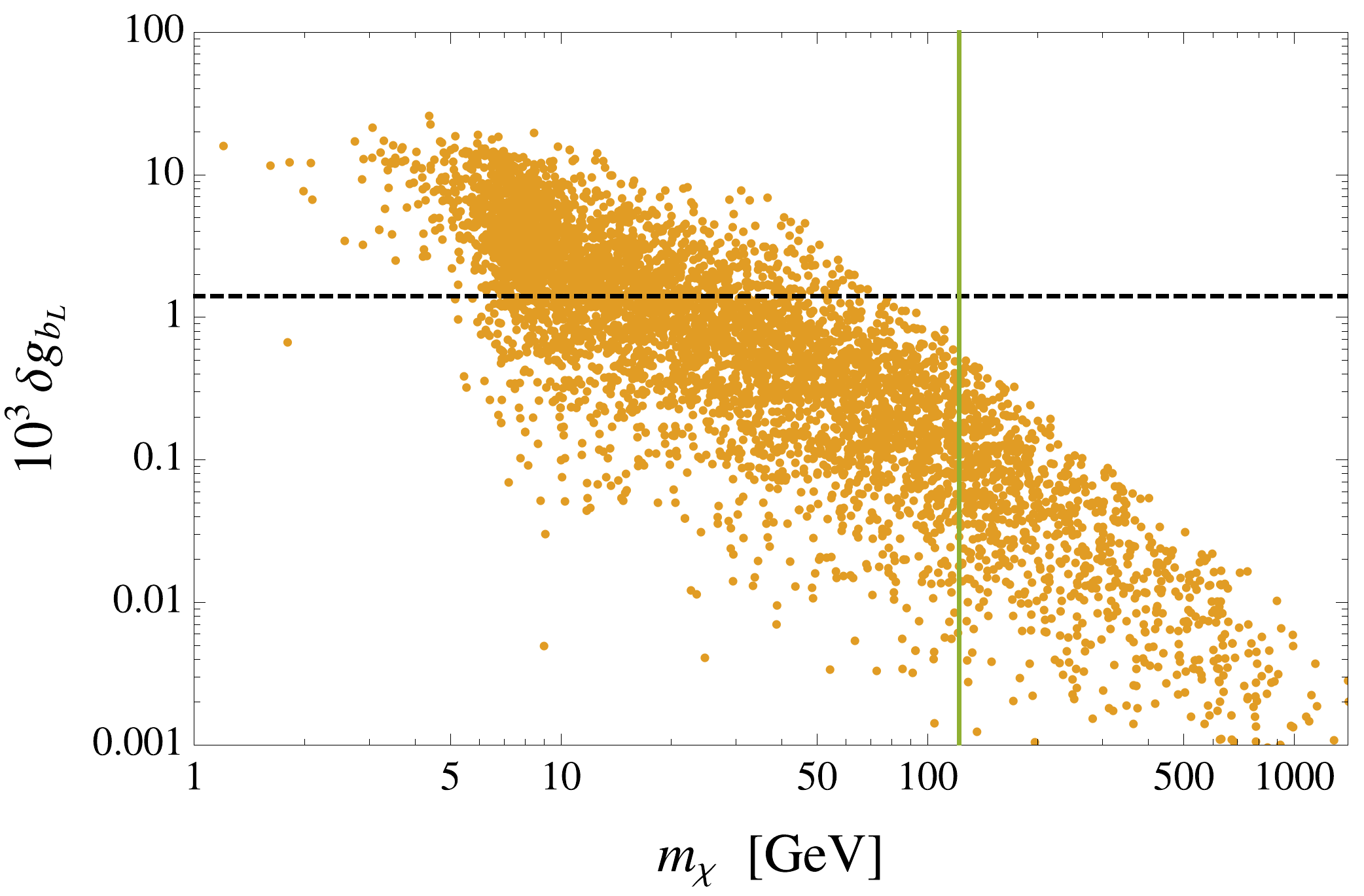}
\caption{Results of the parameter scan of the model where the DM shift symmetry is broken by $b_R$, for $f = 1\;\mathrm{TeV}$. {\it Left panel:} distribution of the mixings for the two chiralities of the bottom quark. The blue (green) curve corresponds to the relation $y_b \simeq \epsilon_L^b \epsilon_R^b M_{\ast b} / f$ with $M_{\ast b} = 8\,(16)\,\mathrm{TeV}$. Notice that $M_{\ast b} = m_{Q^{(b)}} + m_{S^{(b)}}$ is not a physical mass, and can therefore exceed $4\pi f$. {\it Right panel:} tree-level correction to the $Z\bar{b}_L b_L$ coupling versus the physical $\chi$ mass. The black dashed line indicates the $99\%$ CL experimental upper bound, $10^3 \, \delta g_{b_L} < 1.4$, whereas the green vertical line corresponds to the mass for which $\chi$ yields the observed DM density by annihilating purely through the derivative Higgs portal.}
\label{fig:scan}
\end{figure}

For illustration a numerical scan of the model parameter space was performed, setting $f = 1\;\mathrm{TeV}$ and requiring that the scalar potential generated by the top and bottom sectors gives the observed Higgs VEV and mass. We chose $m_{Q^{(b)}} > m_{S^{(b)}}$, yielding $0 < C_{QS} < 1$ and $\mu^2_{\rm DM} > 0$, therefore $U(1)_{\rm DM}$ is never spontaneously broken. In addition, we took $\Lambda = 10 f$ as UV cutoff for the bottom contributions to $\mu^2_{\rm DM}$ and $\mu^2_h$. The results of the scan are reported in Fig.~\ref{fig:scan}, where to approximately account for LHC constraints~\cite{Aaboud:2017zfn} only points where all resonances are heavier than $1.2\;\mathrm{TeV}$ are shown. The distribution of the mixings $\epsilon_{L,R}^b$, shown in the left panel, clearly follows Eq.~\eqref{eq:yukawa} and significantly populates the region $\epsilon_L^b \sim \epsilon_R^b \sim \sqrt{y_b f/ M_{\ast b}} \sim 0.03\,$-$\,0.04$, where the parametric scalings in Eq.~\eqref{eq:parametersbBreaking} approximately apply.
From the right panel, which shows the tree-level $\delta g_{b_L}$ versus the physical $\chi$ mass, we read that in the region where $\chi$ constitutes all or part of the observed DM, i.e. $m_\chi \geq m_{\chi}^{(f \,=\, 1\;\mathrm{TeV})} \approx 122\;\mathrm{GeV}$, the tree-level correction to $Z \bar{b}_L b_L$ is always below the experimental bound.

{\bf DM shift symmetry preserved by $b$ quark} The right-handed bottom is embedded as
\begin{equation}
\mathbf{21}_{2/3} \sim \xi_R^{(b)} = \frac{b_R}{2\sqrt{2}}\, \begin{pmatrix}  \mathbf{0}_{2\times 2} &  \begin{matrix} 1 & i \\  -i & 1 \end{matrix} & \\  \begin{matrix} -1 & i \\  -i & -1 \end{matrix} & \mathbf{0}_{2\times 2}  \\  & &   \mathbf{0}_{3\times 3} \end{pmatrix} ,
\end{equation}
where empty entries are zeros. Therefore the embedding of $q_L$ in Eq.~\eqref{eq:TopEmbeddings} is sufficient to generate the bottom mass, and an $X = -1/3$ sector needs not be introduced. The Lagrangian that mixes the $b_R$ with the composite resonances reads
\begin{equation}
\mathcal{\widetilde{L}}_{\rm mix}^{(b)} =\, \epsilon^j_{bQ} \bar{\xi}_R^{(b) BA}U_{Aa}   U_{B7} Q^{a}_{L, j} + \epsilon^k_{bG} \bar{\xi}_R^{(b) BA} U_{Aa} U_{Bb} G^{ab}_{L, k} + \mathrm{h.c.}, 
\end{equation}
and the complete fermionic Lagrangian is $\mathcal{L}_f = (\mathrm{kin.\;terms}) + (\mathrm{resonance\;masses}) + \mathcal{L}_{\rm mix}^{(t)} +  \mathcal{\widetilde{L}}_{\rm mix}^{(b)} $.

\vspace{1cm}
As a final remark, we have neglected one-derivative operators built out of fermionic resonances, such as (schematically) $\bar{S}_i \slashed{d}^{\,a} Q_{j}^{a}$ and $\bar{Q}_j^a \slashed{d}^{\,b} G_k^{ab}$ in the $X = 2/3$ sector and $\bar{S}^{(b)}_m \slashed{d}^{\,a}Q_n^{(b) a}$ in the $X = -1/3$ sector, which are generically expected to appear in $\mathcal{L}_f$ with $O(1)$ coefficients. Their presence does not affect our discussion, but can have important effects on the resonance phenomenology at high-energy colliders \cite{DeSimone:2012fs}.

\section{$SO(7)/SO(6)$ model: gauge sector} \label{sec:appB}
The $d_\mu$ and $e_\mu$ symbols are computed starting from $D_\mu U = \partial_\mu U - i A_{\mu}^{\hat{a}} T^{\hat{a}} U$, where
\begin{equation} \label{eq:gaugedAmu}
A_{\mu}^{\hat{a}} T^{\hat{a}} = \bar{g} \bar{W}_\mu^{\alpha} T_{L}^\alpha + \bar{g}^\prime \bar{B}_\mu T_{R}^3 + \sqrt{2}\,\bar{g}_D \bar{A}_{D \mu} T^{\rm DM}
\end{equation}
when $SU(2)_L \times U(1)_Y \times U(1)_{\rm DM} \subset SO(6)$ are gauged. The normalization of the $U(1)_{\rm DM}$ gauge coupling is chosen in such a way that the $\chi$ kinetic term obtained from the two-derivative GB Lagrangian
\begin{equation} \label{eq:2deriv}
\mathcal{L}_\pi = \frac{f^2}{4}\, d_\mu^a d^{a\,\mu} 
\end{equation}
is $| (\partial_\mu - i \bar{g}_D \bar{A}_{D\mu}) \chi|^2$. The Lagrangian describing the $SO(6)$ resonance multiplets $\rho_\mu \equiv \rho_\mu^{\hat{a}} t^{\hat{a}} \sim \mathbf{15}$ and $a_\mu \equiv a_\mu^a X^a \sim \mathbf{6}$ was given in Eq.~(A12) of Ref.~\cite{Balkin:2017aep}. $\rho_\mu$ contains an $SO(4) \times U(1)_{\rm DM}$ singlet $\rho_D$ that mixes with $\bar{A}_D$. The mass matrix and the rotation that diagonalizes it are
\begin{equation} \label{eq:MassMatrixGauge}
\frac{f_{\rho}^2}{2}\,  (\bar{A}_D, \rho_D) \begin{pmatrix} 2\bar{g}_D^2 & - \sqrt{2}\bar{g}_D g_\rho \\ - \sqrt{2}\bar{g}_D g_\rho & g_\rho^2  \end{pmatrix}  \begin{pmatrix} \bar{A}_D \\ \rho_D \end{pmatrix}\,, \; \begin{pmatrix} \bar{A}_D \\ \rho_D \end{pmatrix} \to \begin{pmatrix} \tfrac{g_\rho}{\sqrt{g_\rho^2 + 2 \bar{g}_D^2 }} & - \tfrac{\sqrt{2} \bar{g}_D}{\sqrt{g_\rho^2 + 2 \bar{g}_D^2}} \\  \tfrac{\sqrt{2} \bar{g}_D}{\sqrt{g_\rho^2 + 2 \bar{g}_D^2}} & \tfrac{g_\rho}{\sqrt{g_\rho^2 + 2 \bar{g}_D^2 }} \end{pmatrix} \begin{pmatrix} A_D \\ \rho_D \end{pmatrix} ,
\end{equation}
hence the physical dark photon coupling is $g_D = g_\rho \bar{g}_D / \sqrt{g_\rho^2 + 2 \bar{g}_D^2}\,$.

Integrating out the vector resonances at tree level we obtain an effective Lagrangian for the elementary gauge fields and the GBs, $\mathcal{L}_g^{\rm eff} + \delta \mathcal{L}_g^{\rm eff}$, where $\mathcal{L}_g^{\rm eff}$ was given in Eq.~(B1) of Ref.~\cite{Balkin:2017aep}, while
\begin{equation} \label{eq:LeffLandauGauge}
\delta \mathcal{L}_g^{\rm eff} = \frac{1}{2} \Big( g^{\mu \nu} - \frac{p^\mu p^\nu}{p^2} \Big) \Pi_{A A} \bar{A}_{D \mu} \bar{A}_{D \nu} , \qquad   \Pi_{AA}  = \Pi_{A} + \frac{2\bar{g}_D^2}{\bar{g}^2} \frac{\chi^\ast \chi}{f^2}\, \Pi_1^g ,
\end{equation}
with Euclidean-space form factors
\begin{equation}
\Pi_{A} = p^2 \left( 1 + \frac{2 \bar{g}_D^2 f_\rho^2}{p^2 + m_\rho^2} \right), \qquad \Pi_1^g = \bar{g}^2 \left[f^2 + 2 p^2 \left( \frac{f_a^2}{p^2 + m_a^2} - \frac{f_\rho^2}{p^2 + m_\rho^2} \right) \right].
\end{equation}
The one-loop effective potential is $V_g (\tilde{h}) + \delta V_g (\chi)$, where the Higgs-dependent piece was given in Eq.~(B4) of Ref.~\cite{Balkin:2017aep}, and 
\begin{equation} \label{eq:deltaVg}
\delta V_g (\chi) = \frac{3}{2} \int \frac{d^4 p}{(2\pi)^4} \log \left[ 1 + \frac{\bar{g}_D^2}{\bar{g}^2} \frac{2\chi^\ast \chi}{f^2} \frac{\Pi_1^g}{\Pi_A} \right].
\end{equation}
Note that, importantly, the one-loop potential does not contain a Higgs portal term $\sim \lambda \tilde{h}^2 \chi^\ast \chi$. Expanding the logarithm in Eq.~\eqref{eq:deltaVg} and matching to Eq.~\eqref{eq:Veff} gives for the $\chi$ mass term
\begin{equation} \label{eq:DMmassRadiative}
\mu^2_{\rm DM} = \frac{3 \bar{g}_D^2}{16\pi^2 \bar{g}^2 f^2} \int_0^\infty dp^2 p^2 \frac{\Pi_1^g}{\Pi_A}\,,
\end{equation}
which is in general quadratically UV-divergent, but is automatically rendered finite after the two WSRs that ensure finiteness of $V_g (\tilde{h})$ are imposed, namely $2 (f_\rho^2 - f_a^2) = f^2$ and $f_\rho^2 m_\rho^2 = f_a^2 m_a^2$ \cite{Balkin:2017aep}. After the WSRs are used to express $f_a, m_a$ in terms of $f_\rho, m_\rho$, and $f$, we find $\Pi_1^g > 0$, which guarantees that $U(1)_{\rm DM}$ is never spontaneously broken. Performing the integral and taking the leading order in $\bar{g}_D^2 f^2/m_\rho^2, \bar{g}^2_D f_\rho^2 / m_\rho^2 \ll 1$ we arrive at 
\begin{equation} \label{eq:gaugeDMmass}
\mu^2_{\rm DM} \simeq \frac{3\alpha_D}{2\pi} \frac{f_\rho^2}{f^2}\,m_\rho^2 \log \left( \frac{2 f_\rho^2/f^2}{2 f_\rho^2/f^2 - 1} \right).
\end{equation}
The results above assume massless dark photon. A St\"uckelberg mass can be obtained by extending the coset to $SO(7)\times U(1)^\prime / SO(6)$ and gauging the diagonal combination of $U(1)_{\rm DM} \times U(1)^\prime$, namely $\sqrt{2}\,T^{\rm DM} + Z^\prime$. All SM fields are assumed to be uncharged under $U(1)^\prime$. The extended Goldstone matrix is $U = \exp (i \sqrt{2} \pi^a X^a/f) \exp (i \hat{\pi} Z^\prime / f^\prime)$ and the two-derivative Lagrangian becomes $\mathcal{L}_\pi + (f^{\prime\, 2} / 2) \hat{d}_\mu \hat{d}^\mu$, where $\hat{d}_\mu = - (\partial_\mu \hat{\pi} - \bar{g}_D f^\prime \bar{A}_{D \mu})/f^\prime$ to all orders in $1/f^\prime$. The additional piece is precisely the St\"uckelberg Lagrangian, which gives a mass $m_{A} = \bar{g}_D f^{\prime}$ to $\bar{A}_D$. In the effective Lagrangian of Eq.~\eqref{eq:LeffLandauGauge} we must then replace $\Pi_A \to \Pi_A + m_A^2$, which in turn leads to a suppression of the $\chi$ mass: taking for simplicity $f_\rho = f$, Eq.~\eqref{eq:gaugeDMmass} becomes
\begin{equation}
\frac{\mu_{\rm DM}^2 (m_A^2)}{ \mu_{\rm DM}^2 (0) }\Big|_{f_\rho \,=\,f} = \frac{1 + \tfrac{y}{2(1-y)} \tfrac{\log y}{\log 2}}{1 - \tfrac{y}{2}}\,, \qquad y \equiv \frac{m_A^2}{m_\rho^2}\,.
\end{equation}
Numerically, the suppression is small: for example $\mu^2_{\rm DM} (m_A^2)/\mu^2_{\rm DM} (0) \approx 0.97$ for $m_A / m_\rho = 1/10$. As long as $m_A^2/m_\rho^2,\, \bar{g}_D^2/g_\rho^2 \ll 1$, after $m_A^2$ is included in the mass matrix in Eq.~\eqref{eq:MassMatrixGauge} the diagonalization is still obtained through a rotation of angle $\theta \sim \sqrt{2}\bar{g}_D/g_\rho\,$.

\section{$SO(7)/SO(6)$ model: $U(1)_Y\,$-$\,U(1)_{\rm DM}$ kinetic mixing} \label{app:KinMixing}
In this appendix we show that kinetic mixing of $U(1)_Y$ and $U(1)_{\rm DM}$ (in short, $Y$-DM kinetic mixing) can vanish exactly in the $SO(7)/SO(6)$ model, thus motivating the choice $\varepsilon = 0$ made throughout our discussion. 

As first step, we neglect the explicit $\mathcal{G}$ breaking in the fermion sector and consider the bosonic Lagrangian including the gauging of $SU(2)_L \times U(1)_Y \times U(1)_{\rm DM}\,$. At $O(p^2)$ this is simply given by Eq.~\eqref{eq:2deriv}, and the kinetic mixing operators arise at $O(p^4)$. The four-derivative bosonic Lagrangian was first written down for the $SO(5)/SO(4)$ model in Ref.~\cite{Contino:2011np}. To obtain a basis of operators for our model we find it convenient to follow Ref.~\cite{Alonso:2014wta}, where the $O(p^4)$ Lagrangian for $SO(5)/SO(4)$ was discussed by parametrizing the GBs with the matrix $\Sigma (\vec{\pi}) = U(\vec{\pi})^2$. This alternative, but equivalent, description is possible for {\it symmetric} cosets such as $SO(N+1)/SO(N)$, which admit an automorphism (grading) $\mathcal{R}$ of the algebra that flips the sign of only the broken generators, $T^{\hat{a}} \to +\, T^{\hat{a}}$ and $X^a \to - X^a$. The three building blocks that are used to construct invariant operators, all transforming in the adjoint of $\mathcal{G}$, are
\begin{equation} \label{eq:p4BuildingBlocks}
V_\mu = (D_\mu \Sigma)\Sigma^{-1} , \qquad  A_{\mu \nu} = \partial_\mu A_\nu - \partial_\nu A_\mu - i [A_\mu, A_\nu], \qquad  \Sigma A_{\mu \nu}^{\mathcal{R}} \Sigma^{-1} ,
\end{equation}
where $A_{\mu \nu}^{\mathcal{R}} \equiv \mathcal{R} (A_{\mu\nu})$ and we formally took the whole of $\mathcal{G}$ to be gauged by $A_\mu = g_{\mathcal{G}} A_\mu^A T^A$, hence the covariant derivative is $D_\mu \Sigma = \partial_\mu \Sigma - i (A_\mu \Sigma - \Sigma A_\mu^{\mathcal{R}})$. In this formalism, the two-derivative Lagrangian is $\mathcal{L}_\pi = - (f^2/16) \mathrm{Tr}\big[V_\mu V^\mu\big]$. 

In our model the physical sources are given by Eq.~\eqref{eq:gaugedAmu}, which satisfies $A_\mu^{\mathcal{R}} = A_\mu$. By constructing a complete basis for the $O(p^4)$ Lagrangian $\mathcal{L}_4$, we find that $Y$-DM kinetic mixing is encoded by the operators
\begin{equation} \label{eq:p4kinmix}
\mathrm {Tr} \big[ \mathbf{\bar{B}}_{\mu \nu} \mathbf{\bar{F}}_{D}^{\mu \nu} \big], \qquad \mathrm {Tr} \big[ \Sigma \mathbf{\bar{B}}_{\mu \nu} \Sigma^{-1} \mathbf{\bar{F}}_{D}^{\mu \nu} \big] ,
\end{equation}
where $\mathbf{\bar{B}}^{\mu \nu} \equiv \bar{g}^\prime \bar{B}^{\mu \nu} T_R^3$ and $\mathbf{\bar{F}}_D^{\mu \nu} \equiv \sqrt{2}\, \bar{g}_D \bar{F}^{\mu \nu}_D T^{\rm DM}$. Both operators in Eq.~\eqref{eq:p4kinmix} vanish identically. In fact, we have checked that the whole $\mathcal{L}_\pi + \mathcal{L}_4$ is invariant under the parity $P_6 = \mathrm{diag}\,(1, 1, 1, 1, 1, -1, 1) \in O(7)$. Recalling that $T^{\rm DM}$ generates rotations in the $(5,6)$ plane [$\sqrt{2} \,T^{\rm DM} = \mathrm{diag}\,(\mathbf{0}_{4\times 4}, \sigma^2, 0)$], $P_6$ is identified with the charge conjugation $C_D$ that we referred to in the main text. The action of $P_6$ on the $SO(7)$ generators is
\begin{equation} \label{eq:P6generators}
P_6\, T P_6 = +\, T, \quad T = \left\{ T_{L,R}^\alpha, T^{a5}, X^b\right\} \qquad \mathrm{and} \qquad P_6\, \mathcal{T} P_6 = -\,  \mathcal{T}, \quad  \mathcal{T} = \left\{ T^{\rm DM}, T^{a6}, X^6\right\}
\end{equation}
where $a = 1, \ldots,  4$ and $b = 1, \ldots, 5$. As a consequence, the GBs and the elementary gauge fields transform as
\begin{equation} \label{eq:P6elem}
\chi \to - \chi^\ast , \qquad \bar{A}_D \to - \bar{A}_{D}\,, \qquad \{h_i, \bar{W}, \bar{B} \} \to + \,\{h_i, \bar{W}, \bar{B} \}  \qquad (i =1, \ldots, 4),
\end{equation}
which shows that if $P_6$ is exact, $Y$-DM kinetic mixing is forbidden. Furthermore, ``higher-derivative kinetic mixing'' operators (i.e. operators that mix $\bar{B}^{\mu\nu}$ and $\bar{F}_D^{\mu\nu}$, but with the insertion of additional derivatives) also have to be built out of the objects in Eq.~\eqref{eq:p4BuildingBlocks}, and are found to vanish. Summarizing our results thus far, the explicit breaking of $SO(7)$ due to the weak gauging does not generate $Y$-DM kinetic mixing.

As second step, we turn on the explicit $\mathcal{G}$ breaking in the fermion sector. Since $[T^{\rm DM}, P_6] \neq 0$, the SM fermions cannot be simultaneously assigned a nonzero $U(1)_{\rm DM}$ charge and definite $P_6$ parity. Therefore if the SM fermions were taken to have $Q_{\rm DM} \neq 0$, then fermion loops would generate $Y$-DM kinetic mixing: for example, this would happen if $q_L$ were embedded in the $(\mathbf{2}, \mathbf{2})_{+1} \subset \mathbf{21}_{2/3}$ of $SO(7)\times U(1)_X$ and $t_R$ in the $(\mathbf{1}, \mathbf{1})_{+1} \subset \mathbf{7}_{2/3}$. However, for our purposes we must take $Q_{\rm DM} = 0$ for all SM fields, in order for $\chi$ to be the lightest $U(1)_{\rm DM}$-charged particle and therefore stable. In this case each elementary fermion can be assigned definite parity (all the fermion embeddings employed in this paper have in fact $P_6 = + 1$), which guarantees that fermion loops do not generate $Y$-DM kinetic mixing. 

Note that the last conclusion can be altered by subleading spurions, if a single elementary fermion couples to operators with different $P_6$. As a concrete example we can imagine that $t_R$ has, in addition to the embedding in the $(\mathbf{1}, \mathbf{3})_0 \subset \mathbf{21}_{2/3}$ given in Eq.~\eqref{eq:TopEmbeddings}, a second embedding in the $(\mathbf{1}, \mathbf{1})_0 \subset \mathbf{21}_{2/3}$, namely $\xi_R^{\prime (t)} = t_R T^{\rm DM}$. Then it is clear from Eq.~\eqref{eq:P6generators} that the first spurion has $P_6 = + 1$ while the second has $P_6 = -1$, so $t_R$ cannot be assigned a definite parity. Nonetheless, $P_6$ invariance of the fermionic Lagrangian can still be {\it enforced}, by imposing that each elementary field couples to only even operators (or only odd ones, although we are not interested in that possibility here).

Notice that from Eq.~\eqref{eq:P6generators} it follows that $P_6$ also acts on the resonances: taking as examples the $S$, $Q$ and $G$ fermionic multiplets, we have
\begin{equation}
\mathcal{Y} \leftrightarrow - \mathcal{Z}\,, \qquad  \widetilde{S} \to - \widetilde{S},\qquad \{ \mathcal{T}^{(+)} , \mathcal{B}^{(+)} , \mathcal{X}_{2/3}^{(+)} , \mathcal{X}_{5/3}^{(+)}  \} \leftrightarrow -\, \{ \mathcal{T}^{(-)} , \mathcal{B}^{(-)} , \mathcal{X}_{2/3}^{(-)} , \mathcal{X}_{5/3}^{(-)} \},
\end{equation}
while all the other components are left invariant. One can similarly derive the transformation properties of the other fermionic resonances and of the vector multiplets, where in particular $\rho_D \to - \rho_D$.

\section{Collected results for phenomenology} \label{app:PhenoResults}
The spin-independent DM-nucleon cross section is approximately given by
\begin{equation}
\sigma_{\rm SI}^{\chi N} \approx \frac{1}{\pi} \frac{m_N^4}{m_\chi^2} \left[\,\sum_{q = u, d, s} f_{T_q} a_q + \frac{2}{27} f_{T_g} \Big( \sum_{q = c , b, t} a_q \Big) \right]^2 ,
\end{equation}
where we assumed $m_\chi \gg m_N$ with $m_N$ the average nucleon mass, and the form factors averaged over proton and neutron take the values $f_{T_{u, d}} \approx 0.020, 0.043$ \cite{Hoferichter:2015dsa,RuizdeElvira:2017stg} (see also Ref.~\cite{Alarcon:2011zs}), $f_{T_s} = 0.043$ \cite{Junnarkar:2013ac} (also Ref.~\cite{Ren:2014vea}) and $f_{T_g} \approx 0.89$. See Appendix~C of Ref.~\cite{Balkin:2017aep} for a more detailed description. The expression of the coefficients $a_q$ was given in Eq.~\eqref{eq:ddGEN}. In the scenario where the $\chi$ shift symmetry is broken by $b_R$, $\lambda$ is negligible and the cross section takes the form in Eq.~\eqref{eq:DDxsectionBottom}, where $\tilde{f}_N = (2/27) f_{T_g} \approx 0.066$ in case I and $\tilde{f}_N = f_{T_d} + f_{T_s} + (2/27) f_{T_g} \approx 0.15$ in case II. 

The loop function for the $h\to \gamma_D \gamma_D$ decay is, for $m_{\gamma_D} = 0 $,
\begin{equation} \label{eq:hgaDgaD}
F(\tau) = \frac{\tau}{3} A_0 (\tau),\quad A_0(\tau) = \frac{3}{\tau^2} [f(\tau) - \tau]\,, \;\;\; f(\tau) = \left\{
     \begin{array}{lr}
       \arcsin^{2}\sqrt{\tau}\,, & \quad \tau\leq 1\,,\\
       -\frac{1}{4}\left[\log\left(\frac{1+\sqrt{1-1/\tau}}{1-\sqrt{1-1/\tau}}\right)-i\pi\right]^{2}\,, & \;\;\,\, \tau > 1\,.
     \end{array}
   \right.
\end{equation}
Note that $A_0 (\tau) = 1 + O(\tau)$ for small $\tau$. 

Finally we report the thermally averaged cross sections relevant to the region $m_\chi \lesssim m_{\gamma_D} < 2 m_\chi$. The one for $\chi \chi^\ast \to \gamma_D \gamma_D$ is
\begin{equation} \label{eq:annihilationDarkPhotonMassive}
\langle \sigma_{\chi \chi^\ast \to \gamma_D \gamma_D} v_{\rm rel} \rangle = \frac{2\pi \alpha_D^2}{m_\chi^2}\, \sqrt{1 - R}\; \frac{1 - R + 3 R^2/8 } {(1 - R/2 )^2 }\,, \qquad R \equiv \frac{m_{\gamma_D}^2}{m_\chi^2}\,,
\end{equation}
whereas
\begin{equation} \label{eq:annihilationChiChi}
\langle \sigma_{\gamma_D \gamma_D \to \chi \chi^\ast} v_{\rm rel} \rangle = \frac{22\pi \alpha_D^2}{9m_{\gamma_D}^2}\, (1 - \widetilde{R})^{1/2}\; \Big(1 - \frac{24 \widetilde{R}}{11} + \frac{16 \widetilde{R}^2}{11} \Big)\,, \qquad \widetilde{R} \equiv \frac{m_\chi^2}{m_{\gamma_D}^2}\,.
\end{equation}
For the semi-annihilation
\begin{equation} \label{eq:semiannihilationGammah}
\langle \sigma_{\gamma_D h \to \chi \chi^\ast } v_{\rm rel} \rangle = \frac{\alpha_D v^2 m_h (m_{\gamma_D} + m_h)^2}{6 f^4 m_{\gamma_D}^3}\, \Big[ 1 - \frac{4 m_\chi^2}{(m_{\gamma_D} + m_h)^2} \Big]^{3/2} ,
\end{equation}
while for the inverse process we have
\begin{equation} \label{eq:semiannihilationChiChi}
\langle \sigma_{\chi \chi^\ast \to \gamma_D h} v_{\rm rel} \rangle = \frac{\alpha_D v^2 m_h^4 T}{8 f^4 m_\chi^5}\, \beta_{h \gamma_D}\; \frac{1 - (\tfrac{3}{2}R_{\gamma_D} + 2 R_h) + R_h (R_h + 3 R_{\gamma_D}) + \tfrac{1}{2} R_{\gamma_D} (R_{\gamma_D} - R_h)^2}{(1 - R_h - R_{\gamma_D})^4} \,,
\end{equation}
where $\beta_{h \gamma_D} \equiv [1 + (R_h - R_{\gamma_D})^2 - 2 (R_h + R_{\gamma_D})]^{1/2}$ and $R_i \equiv m_i^2 / (4 m_{\chi}^2)$. Notice the additional factor $T/m_\chi$ coming from the $p$-wave suppression. 

Lastly,
\begin{align} 
\langle \sigma_{\gamma_D \chi \to h \chi } v_{\rm rel} \rangle =\,& \frac{\alpha_D v^2 m_h^4 \,G(m_{\gamma_D}, m_h; m_\chi)}{24 f^4 m_\chi m_{\gamma_D}^3  } \, , \quad \langle \sigma_{h \chi \to \gamma_D \chi } v_{\rm rel} \rangle = \frac{\alpha_D v^2 m_h m_{\gamma_D}^2 G(m_h, m_{\gamma_D} ; m_\chi)}{8 f^4 m_\chi (m_h + 2m_\chi)^2 }\,, \nonumber \\
&\,G(m_1, m_2 ; m_\chi) = \, \frac{ \big[( m_1^2 - m_2^2 ) ( (m_1 + 2m_{\chi})^2 - m_2^2 ) \big]^{3/2}}{ (m_1 + m_{\chi})^2 [m_{\chi} (m_1 + 2 m_{\chi}) - m_2^2 ]^2 }\,. \label{eq:semiannihilationGammaChi}
\end{align}

%%%%%%%%%%%%
%\bibliographystyle{utphys}
%\bibliography{./}

\end{document}